\documentclass[sn-mathphys-num]{sn-jnl}

\usepackage{graphics,graphicx}
\usepackage{amsmath, amssymb, longtable}
\usepackage{url}
\usepackage{multibib}
\usepackage{hyperref}
\usepackage{color}
\usepackage[T1]{fontenc}
\usepackage{multirow}   

\def\aap{Astron. Astrophys.}                
\def\aj{Astron. J.}

\def\apj{Astrophys. J.} 
\def\apjl{Astrophys. J. Lett.}  
\def\apjs{Astrophys. J. Suppl. Ser.}  
\def\aplett{Astrophys. Lett.}

\def\mnras{Mon. Not. R. Astron. Soc.}

\def\nat{Nature}

\def\procspie{Proc. SPIE}

\def\araa{Annu. Rev. Astron. Astrophys.}
\def\apspr{Astrophys. and Space Phys. Rev.}

\def\ssr{Space Sci. Rev.}

\def\nar{New A Rev.}

\newcommand{\degr}{\hbox{$^{\circ}$}}

\newcommand\fdg{\mbox{$.\!\!^\circ$}}%
\begin{document}

\title{Cygnus X-3 revealed as a Galactic ultraluminous X-ray source by IXPE}

\author*[1,2]{Alexandra Veledina}\email{alexandra.veledina@gmail.com}
\author[3]{Fabio Muleri}
\author[1]{{Juri} {Poutanen}}
\author[4,5,6]{{Jakub} {Podgorn\'{y}}}
\author[5]{{Michal} {Dov{\v c}iak}}
\author[3]{{Fiamma} {Capitanio}}
\author[7,8]{{Eugene} {Churazov}}
\author[3]{{Alessandra} {De Rosa}}
\author[3]{{Alessandro} {Di Marco}}
\author[1]{{Sofia} {Forsblom}} 
% ...

\affil[1]{Department of Physics and Astronomy, FI-20014 University of Turku, Finland}
\affil[2]{Nordita, KTH Royal Institute of Technology and Stockholm University, Hannes Alfv{\'e}ns v{\"a}g 12, SE-106 91, Sweden} 
\affil[3]{INAF Istituto di Astrofisica e Planetologia Spaziali, Via del Fosso del Cavaliere 100, 00133 Roma, Italy}
\affil[4]{Universit\'{e} de Strasbourg, CNRS, Observatoire Astronomique de Strasbourg, UMR 7550, 67000 Strasbourg, France}
\affil[5]{Astronomical Institute of the Czech Academy of Sciences, Bo\v{c}n\'{\i} II 1401/1, 14100 Praha 4, Czech Republic}
\affil[6]{Astronomical Institute, Charles University, V Hole\v{s}ovi\v{c}k\'{a}ch 2, 18000 Praha, Czech Republic}
\affil[7]{Max Planck Institute for Astrophysics, Karl-Schwarzschild-Str 1, D-85741 Garching, Germany}
\affil[8]{Space Research Institute of the Russian Academy of Sciences, Profsoyuznaya Str. 84/32, Moscow 117997, Russia}
% ...

\abstract{
The accretion of matter by compact objects can be inhibited by radiation pressure if the luminosity exceeds the critical value, known as the Eddington limit.
Discovery of ultraluminous X-ray sources has shown that accretion can proceed even when the apparent luminosity significantly exceeds this limit.
High apparent luminosity might be produced thanks to geometric beaming of the radiation by an outflow.
The outflow half-opening angle, which determines the amplification due to beaming, has never been robustly constrained.
Using the Imaging X-ray Polarimetry Explorer, we made the measurement of X-ray polarization in the Galactic X-ray binary Cyg X-3. 
We find high, over 20\%, nearly energy-independent linear polarization, orthogonal to the direction of the radio ejections.
These properties unambiguously indicate the presence of a collimating outflow in the X-ray binary Cyg~X-3 and constrain its half-opening angle, $\lesssim$15$\degr$.
Thus, the source can be used as a laboratory for studying the super-critical accretion regime.
This finding underscores the importance of X-ray polarimetry in advancing our understanding of accreting sources.
}

\maketitle

\section*{Main text}

Cyg X-3 is one of the first sources discovered in the X-ray sky \cite{Giacconi1967}.
It is the brightest X-ray binary in radio wavelengths \cite{Gregory1972,McCollough1999,Corbel2012}, with peak fluxes reaching 20~Jy, and one of the few X-ray binaries where $\gamma$-ray emission has been detected \cite{Atwood2009,Tavani2009}.
Cyg X-3 is also exceptional from the point of view of population synthesis and evolutionary studies \cite{Lommen2005,Belczynski2013}. 
It is the only known Galactic source containing a compact object in a binary orbit with a Wolf-Rayet (WR) star -- an evolved massive star that is characterised by a hydrogen-depleted spectrum \cite{vanKerkwijk1992,vanKerkwijk1996}; it is also the progenitor of a double-degenerate system \cite{Belczynski2013} that will become a source of gravitational wave emission in the distant future.

The optical counterpart is not visible because of the high absorption along the line of sight: the source is located in the Galactic plane at a distance $D=9.67^{+0.53}_{-0.48}$~kpc \cite{ReidMiller-Jones2023,McCollough2016}).
The system parameters have been constrained based on radio, X-ray and infrared properties.
Spatially resolved discrete radio ejections \cite{Marti2000,Miller-Jones2004} are aligned in the north-south direction.
Moreover, the position angle of the intrinsic infrared (IR) polarization (which may be associated with scattering off the circumstellar disc \cite{Jones1994} or with the jet \cite{Fender1999}) agrees, within uncertainties, with the jet position angle.
The orbital period $P_{\rm orb}=4.79234^{\rm h}$ has been measured with high accuracy based on the prominent X-ray and IR flux modulations, as well as from the periodic Doppler shifts of the X-ray and IR lines \cite{Hjellming1973,Fender1999,Vilhu2009,Kallman2019}, and is known to change rapidly over time \cite{vanderKlis1981,Antokhin2019}.
Orbital variations are clearly pronounced in all wavelengths, from $\gamma$-rays to radio, and enable determination of the orbital inclination (see Methods and \ref{fig:pol_vs_time}, \ref{fig:xray_lc} and \ref{fig:radio_lc}).
Recent comprehensive analysis of the orbital photometric variations in X-rays and IR  \cite{Antokhin2022} gave the most precise orbital inclination of the source, $i=29\fdg5\pm1\fdg2$. 
Estimates obtained using other methods, Doppler shifts of X-ray lines \cite{Vilhu2009} and relativistic ejections \cite{Mioduszewski2001,Miller-Jones2004}, give consistently small inclinations.

Cyg X-3 swings between several X-ray spectral states, which are tightly linked to radio properties (\ref{fig:diagram} and \ref{fig:broadband_spectra}) \cite{Szostek2008_states}. 
It spends most of the time in the hard X-ray, quiescent radio state. 
In this state, the X-ray emission can be described by a power law with prominent fluorescent iron lines (\ref{fig:nicer_orbit_folded}, see  Methods).
Occasionally, Cyg X-3 shows transitions to an ultrasoft spectral state, during which the spectrum is dominated by a blackbody peaking at a few keV.
Transitions to this state are accompanied by major radio ejections, in which the highest observed radio fluxes are reached.
The spectral transitions are thought to be related to changes of accretion geometry, however, the exact geometrical configuration in each state and the physical reasons behind the changes are not known.

Understanding the physical picture of the system is complicated by the diversity of models that can explain the X-ray spectra. 
The quiescent-state spectra (corresponding to a hard X-ray continuum) can be well fit with either (i) an intrinsically soft spectrum severely absorbed in the WR wind, or (ii) with a hard spectrum coming from the hot medium located within the truncated cold accretion disc (this model is often discussed in the context of other hard-state sources), or (iii) with the equal contribution of the incident spectrum and the reflected emission \cite{Zdziarski2010,Hjalmarsdotter2008}.
Early works identified the potential importance of scattering of intrinsic emission in the formation of the observed spectra \cite{MilgromPines1978,WhiteHolt1982}.
The models invoke very different emission mechanisms and a wide range of inherent luminosities and accretion rates, preventing us from identifying the accretion-ejection mechanisms of this unusual binary.
The astronomical puzzle called Cyg~X-3 \cite{Hjellming1973} remained unsolved for over 50 years after its discovery, even though the system is one of the most frequently studied sources in the X-ray sky.

\begin{figure}[h]
\begin{center}
    \includegraphics[width=0.7\textwidth]{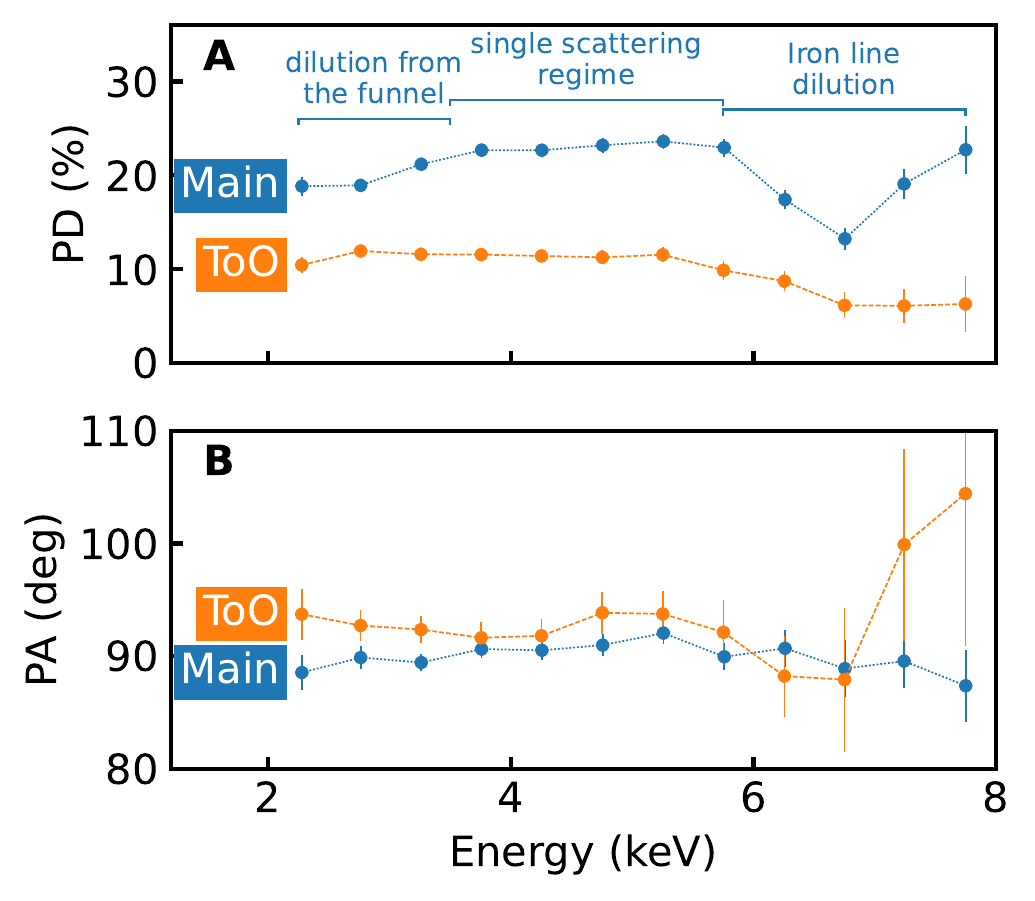}
\end{center}
    \caption{\textbf{Orbital-phase averaged polarization properties from IXPE observations.}
    The energy dependence of the average PD (A) and PA (B) are shown for the Main (blue lines) and ToO (orange lines) observations. In both cases the PA is consistent with constant across the energies and the PD in the 3.5--6~keV band is constant (null hypothesis probability values are 91\% and 78\% for the Main and ToO observations, respectively). A decrease of PD above 6~keV is caused by the contribution of the unpolarized iron line(s). PD is also lower below 3~keV for the Main observation, which may be related to the unpolarized contribution of reprocessed emission by the funnel walls (see Methods). Data are given as the mean values in the selected energy band and the error bars correspond to 1$\sigma$ confidence level.
     \label{fig:qu_ave}}
\end{figure}

We report here on the detection of the X-ray polarization from \mbox{Cyg X-3}.
Observations with the Imaging X-ray Polarimetry Explorer (IXPE) \cite{Weisskopf2022} revealed the accretion-ejection geometry of the source.
The first IXPE observation (hereafter referred to as ``Main'') caught the source in the hard X-ray (radio-quiescent) state and consisted of two runs, 14--19 October 2022 and 31 October--6 November 2022 (Table~\ref{tab:mw_xray}).
We find a high polarization degree PD=$20.6\pm 0.3\%$ in the 2--8\,keV range (see Fig.~\ref{fig:qu_ave}).
The polarization angle PA=$90\fdg1\pm0\fdg4$ (that is determined by the direction of electric field oscillations, measured from north through east on the sky) is orthogonal to the position angle of the discrete radio ejections and the infrared and sub-mm polarization (Table~\ref{tab:mw_radio}) \cite{Jones1994,Marti2000,Miller-Jones2004}.
The observed PD is constant over the 3.5--6~keV range. 
The PD is lower in the 6--8\,keV range, where the fluorescent and recombination Fe K$\alpha$ emission lines dominate, as well as below 3~keV (see Fig.~\ref{fig:qu_ave}A and Methods).
The fluorescent lines originate from the transitions of the electrons to lower atomic levels of iron, and the distribution of the resulting photons and their electric vectors is expected to be isotropic, giving unpolarized emission.

\begin{figure}[h]
\begin{center}
    \includegraphics[width=0.7\textwidth]{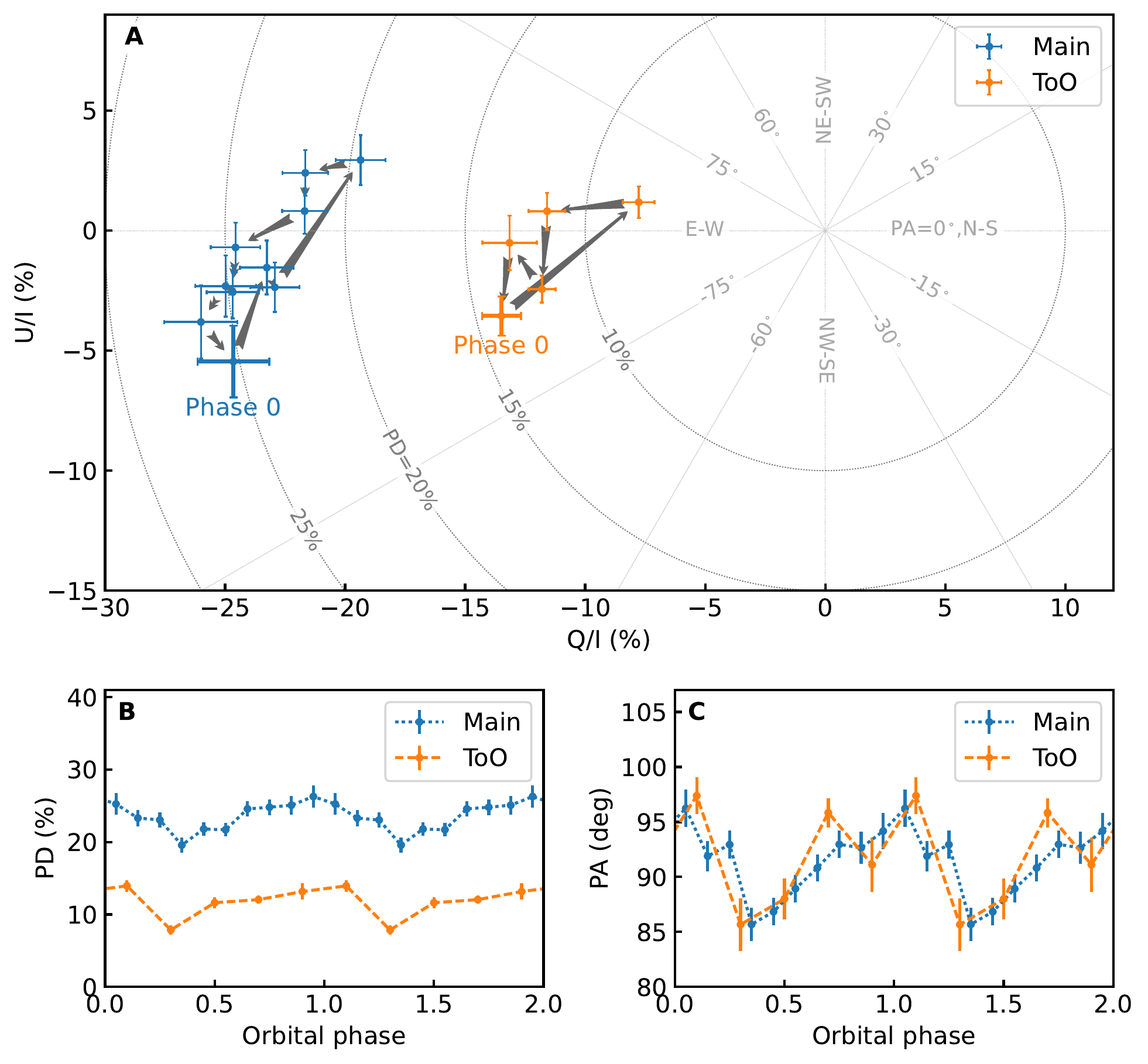}
\end{center}
    \caption{\textbf{Orbital phase-folded polarization properties.}
    (A) Evolution of the normalised Stokes parameters $Q/I$--$U/I$. Blue colour corresponds to the Main observation and orange corresponds to the ToO observation. Arrows indicate the path followed by the data. 
    Phase 0 corresponds to the superior conjunction.
The dependence of the PD (B) and PA (C) on the orbital phase in the 3.5--6~keV energy range. 
Data are presented as mean values in the selected orbital phase bin and the error bars correspond to 1$\sigma$ confidence level.
     \label{fig:qu_orb}}
\end{figure}

We performed an orbital phase-resolved analysis of the polarimetric data using the recent ephemeris (see Methods).
We note large variations of the PA (Fig.~\ref{fig:qu_orb}) and a complex relation between the PA and PD variations (\ref{fig:pol_orb_energy_all} and \ref{fig:bow_shock}, Methods).
The pattern is not consistent with a model of scattering off optically-thin plasma in an orbit with the X-ray source \cite{BME1978}, e.g. scattering off the wind close to the surface of the WR star.
In this case, the low inclination of the system would lead to a sinusoidal variations of PA with two peaks per orbital period (equivalent to a double loop in the normalised Stokes parameters $Q/I$--$U/I$ plane).
Furthermore, the PD of the primary X-rays reflected off the star is expected to be $<$1\%: for a distant reflector, the PD is small due to the small solid angle subtended by the star as seen from the compact object.
For a higher solid angle of the scattering matter, namely if scattering proceeds within the WR wind, a low PD is also expected, as in this case the scatterers are nearly spherically symmetric.

\begin{figure}[h]
\centering
    \includegraphics[width=0.95\textwidth]{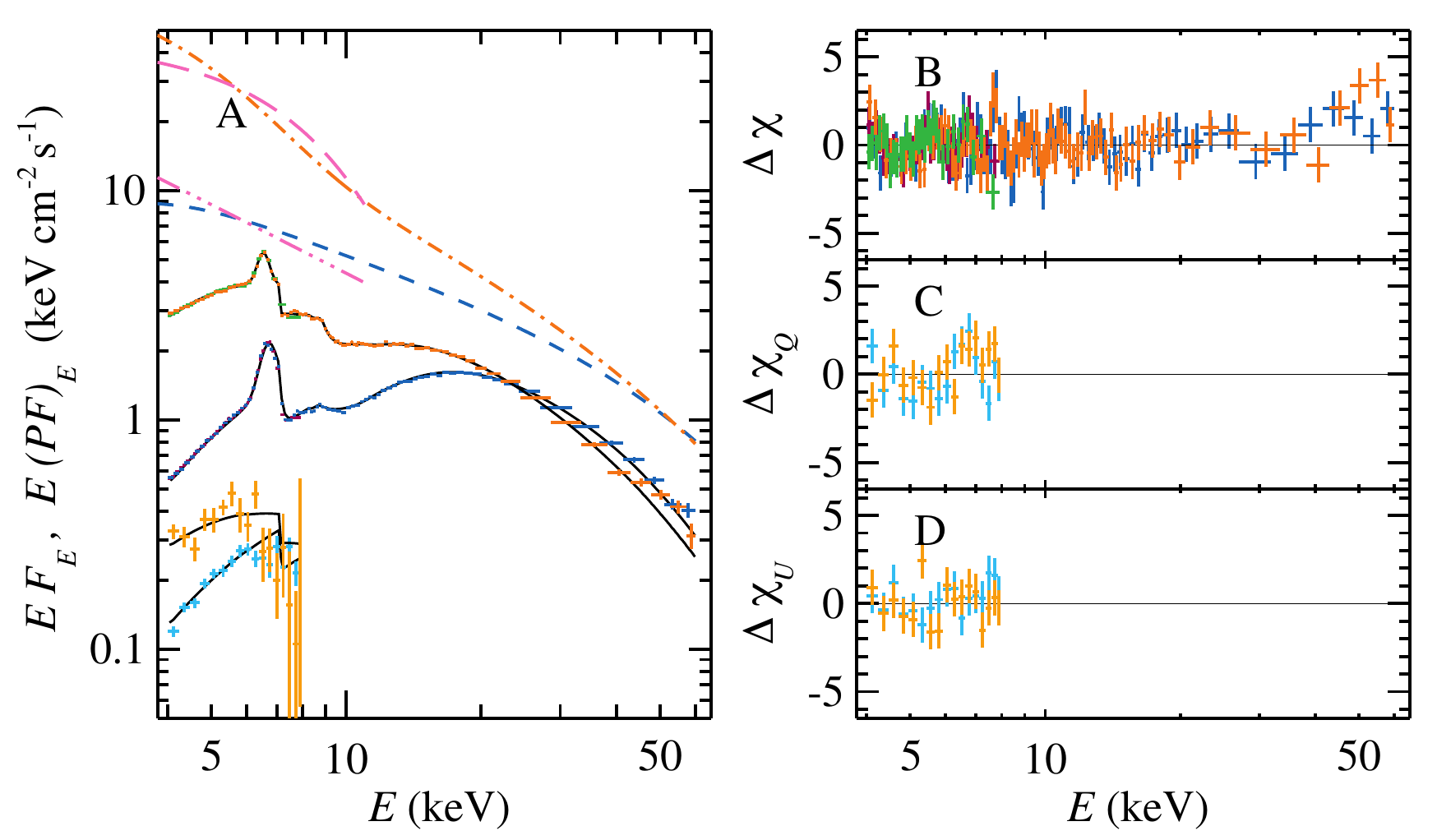}
\caption{\textbf{Average spectropolarimetric data with the best-fitting models for the Main and ToO observations.} 
(A) The observed fluxes in $EF_E$ units: NuSTAR Main (blue) and ToO (orange) and IXPE Main (violet) and ToO (green) observations. Note that IXPE and NuSTAR points are nearly identical and hence may not be well distinguished in the plot. IXPE polarized fluxes $E (PF)_E={\rm PD}(E) EF_E$ are shown for the Main (cyan) and ToO (yellow) observations.
Black lines correspond to the fitting model for each dataset.
Blue dashed and orange dot-dashed lines correspond to the intrinsic spectra that are needed to reproduce the observed reflected spectra, for the Main and ToO datasets, respectively.
Magenta long-dashed and triple-dot-dashed lines correspond to the illustrative intrinsic spectra of supersoft and soft ULXs (spectra and notations are adopted from \cite{Kaaret2017}), overlaid on top of the intrinsic spectra of Cyg X-3 for comparison.
(B), (C) and (D) Residuals of the data relative to the model, in units of the errors, for the fluxes, $Q$ and $U$ parameters, respectively.
The error bars correspond to 1$\sigma$ confidence level.
}\label{fig:specpol_fits}
\end{figure}

The high, in excess of 20\%, average PD and its orientation relative to the radio outflows suggest that the IXPE signal is dominated by the reflected component, with minor to zero contribution of the primary continuum. 
Indeed, the Comptonization continuum cannot give such high polarization, while the synchrotron mechanism fails in explaining the multiwavelength polarization properties: the PA of sub-mm and IR polarization are orthogonal to the X-ray polarization.
This conclusion is bolstered by our finding of a largely energy-independent polarization  as the superposition of comparable contributions of primary and reflected emission would lead to a strong energy dependence of the PD.
Previous spectral modelling included scenarios with substantial contribution of reflection to the total X-ray spectra \cite{Hjalmarsdotter2008}, however, the 2--8~keV spectra have never been considered to be completely dominated by reflection.
To verify the possibility of the dominant contribution of the reflection component in the IXPE band, we performed spectro-polarimetric modelling (Fig.~\ref{fig:specpol_fits} and Table~\ref{tab:specpol}).
We used the model comprised of the reflection continuum and a Gaussian component that mimics contribution of the iron line complex at energies 6--7~keV.
PDs of both components and PA of the continuum are free fitting parameters, and the PA of the Gaussian is fixed to the value found for the continuum (see Methods). 
We find that the observed broadband spectral energy distribution (SED) can be well approximated ($\chi^2$/d.o.f.=936/881) with reflection of an intrinsically soft spectrum.
The reflection continuum is highly polarized with constant PD=$22.8\pm 0.4\%$ and the line emission is unpolarized (with a 90\% confidence level upper limit of 2\%).

\begin{figure}[h]
\begin{center}
    \includegraphics[width=\textwidth]{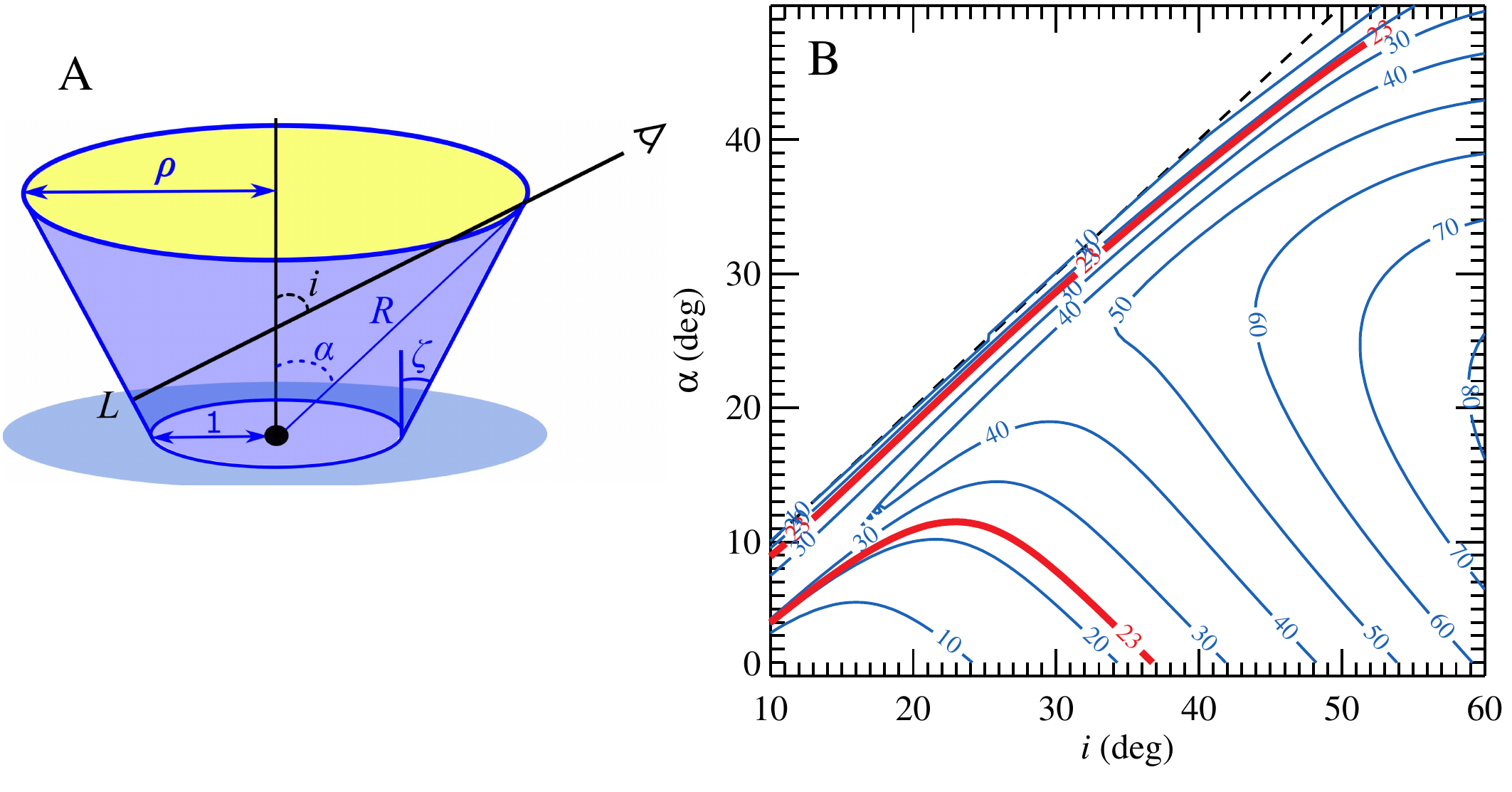}  
\end{center}
    \caption{\textbf{Geometry of the funnel and its polarization properties.}
    (A) Proposed geometry of the funnel with the primary emitting source marked by a filled black circle. The observer located at inclination $i$ sees the reflection from the inner surface of the funnel (down to point $L$). Primary source is obscured by the funnel walls. The half-opening angle of the funnel $\zeta$ is always smaller than the angle $\alpha$, the latter used as a proxy for $\zeta$ in the text. Radii are measured in units of the funnel base. (B) Contour plots of the constant PD for different observer inclinations $i$ and parameters $\alpha$. 
Numbers in the lines denote polarization in per cent. Red contour marks the observed polarization of $23\%$.}
     \label{fig:model_funnel}
\end{figure}

The polarization of Cyg~X-3 resembles closely that of the accreting supermassive black hole in the Circinus galaxy which exhibits a PD of $28\pm7\%$ \cite{Ursini2023}. 
In that source, the primary X-rays are believed to be obscured by a dusty torus with an inclination exceeding that of the host galaxy, $i\sim65\degr$, so that the reflected emission dominates over the direct emission in the IXPE band.
Obscuration of a system with low, $i\sim30\degr$  \cite{Mioduszewski2001,Vilhu2009,Antokhin2022}, orbital inclination is not naturally expected.
Our finding of the high, energy-independent PD leads to an important implication for the accretion geometry of Cyg~X-3: as the observer does not see the primary X-ray source, we infer the presence of an optically thick medium high above the orbital plane, shaped as a funnel (see Fig.~\ref{fig:model_funnel}).
For the Thomson scattering (see equation~\ref{eq:pd_thomson} in Methods), the observed PD translates to the typical scattering angle $\approx38\degr$, which is close to the orbital inclination.
Our modelling of the X-ray polarization indicates a narrow funnel with $\alpha\lesssim$$15\degr$ half-opening angle.
In Fig.~\ref{fig:model_funnel}B we show the contour plots of the PD as a function of $\alpha$ and $i$ (see also \ref{fig:model_alpha_rho} and \ref{fig:MC_geometries}). We identify two branches of solutions (red lines), however, exclude the upper branch based on the dependence of the orbit-average PD on time (see \ref{fig:pol_vs_time} and Methods).

Optically thick and elevated envelopes are hallmarks of super-Eddington accretion rates \cite{SS73,Poutanen07,Sadowski2014}.
We can check this hypothesis by estimating the intrinsic X-ray luminosity of Cyg X-3 (\ref{fig:model_alpha_rho}C).
Assuming that the observed radiation comes from the visible inner part of the funnel, we can relate the reflected luminosity to the intrinsic one through the reflection albedo and the solid angle of the visible part of the funnel (alternatively, the scattering can proceed in the WR wind right above the funnel, but the resulting luminosities are the same, see more details in Methods).
We find that the intrinsic luminosity exceeds the Eddington limit for a neutron star accretor at half-opening angles $\alpha\approx8\degr$, while for $\alpha\approx15\degr$ this limit is exceeded for a black hole of 10 solar masses.
Further, for the small opening angle of the funnel required by the polarimetric data, the apparent luminosity for an observer viewing down the funnel is $L\gtrsim5.5\times10^{39}$~erg~s$^{-1}$ in 2--8~keV range, which puts Cyg X-3 in the class of ULX sources \cite{Kaaret2017}.

With the aim to identify the properties of the accretion geometry that drive the soft-hard-state transitions in the source, we performed an additional IXPE target of opportunity (``ToO'' hereafter) observation as the source transitioned towards the soft state, as indicated by the X-ray and radio fluxes, on 25--29 December 2022 (see Methods).
The ToO revealed a twice lower, largely energy-independent PD=$10.0\pm 0.5\%$ at 2--8~keV. 
While the PA=$90\fdg6\pm1\fdg2$ is similar to the value found in the Main observation (see orange symbols in Figs.~\ref{fig:qu_ave}--\ref{fig:qu_orb} and \ref{fig:specpol_fits}).
In particular, this suggests that we continue seeing the reflected signal in this state, but the funnel parameters have changed, in particular, the decreased polarization may suggest the reflection and reprocessing now operates in some volume of matter around the funnel (within the outflow material), rather than coming solely from its surface.
This is consistent with the outflow becoming more transparent,  which may be related to the drop of the mass accretion rate.
We expect that a further decrease of the matter supply would lead to a collapse of the funnel, revealing the X-ray emission from the inner parts of the accretion disc, accompanied by the drop of its polarization.
The spectrum in this case should closely resemble the one produced by the multicolour disc \cite{SS73}, which is indeed observed in Cyg~X-3 during the so-called ultrasoft state.
Our scenario suggests that this state corresponds to a lower accretion rate, as compared to the hard X-ray/radio quiescent state, even though the source appears brighter.
Following our findings, the whole complex of multiwavelength properties may need to be reconsidered in light of this renewed accretion geometry.

Narrow funnels have been used to explain the high apparent luminosity of the ultraluminous X-ray sources (ULXs) \cite{King2001,King2009,Kaaret2017,Middleton2021}.
However, determination of the opening angles of the base of these ULX outflows directly from the observations has not been possible, as the observer is located close to the axis of the funnel and sees the strongly amplified radiation of the central source.
At the same time, the Galactic super-critical counterparts are seen at high inclinations: $i\approx78\degr$ in the persistent source SS~433 \cite{Fabrika04SS433}, $60\degr\lesssim i \lesssim80\degr$ in the transient V404~Cyg, $i=66\degr\pm2\degr$ in GRS~1915+105 \cite{Done2004,CasaresJonker2014,Motta2017}, $i=72\degr\pm4\degr$ in V4161~Sgr \cite{Revnivtsev2002,MacDonald2014}, so that a thick outflow blocks the inner engine from our line of sight.
The small inclination of Cyg~X-3 system allows one to see the reflected/scattered component from the inner surface of the funnel, and the properties of X-ray polarization enable robust conclusions about the source geometry.

The X-ray polarimetric data probe the accretion geometry in Cyg X-3.
These data have revealed that this famous and long-studied Galactic source has been accreting in the super-Eddington regime.
This discovery opens a new chapter in the study of this exceptional system, and establishes it as an analogue of distant ULXs and super-critical transients, such as tidal disruption events.
We have directly constrained the half-opening angle of the outflow funnel of Cyg~X-3 to be $\lesssim$15$\degr$.

\section*{Methods}

\subsection*{X-ray spectro-polarimetry analysis}\label{sect:meth_ixpe}

The first attempt to detect the linear polarization of the X-rays from Cyg X-3 was made with the {OSO-8} satellite \cite{Long1980}, but the presence of other bright sources in the field of view prevented the authors to reach firm conclusions.
IXPE \cite{Weisskopf2022} observed Cyg~X-3 twice: the first and second observations were named ``Main'' and ``ToO''. 
The Main observation was split in two observing periods close in time:  the first started on 2022-10-14 01:26:33~UTC and ended on 2022-10-19 14:12:56~UTC, and the second was carried out between 2022-10-31 12:50:08~UTC and 2022-11-06 08:42:21 UTC. 
The ToO observation started on 2022-12-25 10:05:17 UTC and ended on 2022-12-29 17:44:22 UTC. 
The duration of the Main and ToO observation is $\sim$538~ks and $\sim$199~ks, respectively.

The analysis of the IXPE data was carried out similarly to other observations (e.g., see \cite{Krawczynski2022}). 
{Level 2} (processed) data were downloaded from the IXPE HEASARC archive.
These data consist of three photon lists, one for each of the IXPE telescopes, and contain for each collected photon the time, position in the sky, as well as the Stokes parameters of the single event. 
The arrival time of the photons were corrected to the Solar system barycenter using the {\tt barycorr} tool from the {\sc ftools} package, included in \textsc{heasoft}~version 6.31, using the Jet Propulsion Laboratory (JPL) Development Ephemeris (DE421) and the International Celestial Reference System (ICRS) reference frame. 
 
The source extraction region with a radius of 90~arcsec was centered on the source position. 
We did not attempt to extract the background from the remaining part of the field of view and subtract it from the source signal, because the background in the IXPE field of view for relatively bright sources like Cyg~X-3 is relatively weak and is dominated by the contamination of the source photons which are focused in the outer wings of the mirror Half Power Diameter (HPD) \cite{DiMarco23}. 
Thus, removing the background in this case mostly removes a few per cent of the source signal.
 
Polarization can be obtained from the IXPE photon list with two approaches.
The first is building the Stokes spectra $I(E)$, $Q(E)$ and $U(E)$, which are calculated by summing the relevant Stokes parameter for all the events in a specific energy bin. 
Such spectra can then be fitted with a  forward-fitting software, associating for each spectral component a certain polarization model \cite{Strohmayer2017}; in our case, we used \textsc{xspec} version 12.13.0 \cite{Arnaud1996}. 
The second approach relies on the use of \textsc{ixpeobssim} package \cite{Baldini2022}, which calculates the Stokes parameters as the sum of the event values in a certain energy, time or angular bin \cite{Kislat2015}. 
The latter approach does not assume any underlying spectral model. 
Data collected from the three IXPE telescopes were analysed separately, applying the appropriate response matrices (unweighted, version 12, in our case) which are available at the HEASARC CALDB and in the \textsc{ixpeobssim} package.

Average polarization in the entire IXPE energy range (2--8~keV) was calculated with the \textsc{pcube} algorithm included in the \textsc{ixpeobssim/xpbin} tool. 
The PD=$20.6\pm0.3\%$ and $10.4\pm0.03\%$ were found for the Main and ToO observations, respectively, with a PA=$90\fdg1\pm0\fdg4$ and PA=$92\fdg6\pm0\fdg7$, measured east of north. 

More detailed study of X-ray polarization and its relation to various spectral components requires proper spectro-polarimetric modelling.
Broadband X-ray spectral coverage of Cyg X-3 during the Main and ToO runs were performed with the Nuclear Spectroscopic Telescope Array (NuSTAR) observatory.
NuSTAR consists of two identical X-ray telescope modules, referred to as FPMA and FPMB \cite{Harrison2013}.
It provides X-ray imaging, spectroscopy and timing in the energy range of 3--79~keV with an angular resolution of 18 arcsec (FWHM) and spectral resolution of 400 eV (FWHM) at 10 keV.
We use two NuSTAR datasets: the first one was carried out on 2022 November 5 (ObsIDs: 90802323004) with the on-source exposure of $\sim16$ ks (during Main observation) and the second one performed on  2022 December 25 (ObsIDs: 90801336002) with $\sim 36$ ks exposure (during ToO observation).
Both observations covered several orbital cycles of the system, which allowed to perform phase-resolved spectroscopy. 
The NuSTAR data were processed with the standard NuSTAR Data Analysis Software (nustardas 4May21 v2.1.1) provided under heasoft v6.29 with the caldb version 20201217. 
Circular 100 arcsec radius regions were used for both source and background spectra extraction.
The source region was centered on the locations of Cyg X-3 and the background region was selected from a sourceless region in the detector image.
All obtained spectra were grouped to have at least 25 counts per bin using the grppha tool.
The final data analysis (timing and spectral) was performed with the heasoft 6.29 software package.

We performed broadband spectral modelling, for both Main and ToO runs, using the data from NuSTAR and IXPE instruments.
We acknowledge a complexity of such modelling in light of the high-amplitude orbital variability.
The small statistical errors of spectra cause the average spectra to be non-representative, as the orbital variations of flux and hardness alter the average spectral shape.
Furthermore, the lower energies contain prominent spectral lines, which are not well resolved in the IXPE band (see \ref{fig:nicer_orbit_folded}).
Mutual contribution of the unpolarized lines may be responsible for the observed decrease of PD below 3~keV, alternatively, the additional contribution of the reprocessed incident X-ray radiation by the funnel walls may be responsible for the visible decrease.
For the aforementioned reasons we add 1\% systematic errors to the data and consider only the data above 4~keV.
While for the Main observation we find that a good fit can be obtained when summing up all spectra (i.e. the spectral shape does not evolve substantially with the orbital phase), for the ToO observation we found that we may only use spectra averaged over orbital phases 0.25--0.75, i.e. close to the inferior conjunction, when the intrabinary absorption is smallest.
As IXPE observed Cyg X-3 in a relatively bright state for a long time, the large collected number of events made evident small systematic difference among the three IXPE telescopes, affecting significantly only the $I$ spectrum. 
To account for them, we introduced a Multiplicative Power Law (MPL) cross-calibration function which reads $ f\times E^{\gamma}$, similarly to what was done for the analysis of the black hole Cyg X-1 observed by IXPE \cite{Krawczynski2022}.
For the Main observation, we also left the gain and offset of the IXPE energy scale free to vary, and the offset for NuSTAR as suggested by \cite{Madsen2022}.

Motivated by the polarization properties, we consider the model where the spectrum is dominated by the reflection component.
We consider the {\sc xspec} model {\sc reflect} describing the reflection spectrum of the neutral matter \cite{MagdziarzZdziarski1995} that is produced by the continuum composed of the multicolour accretion disc, described by {\sc diskpbb} (where the local blackbody temperature has a power-law dependence on distance to the compact object and the power index is a free model parameter), and a soft Comptonization continuum, {\sc nthComp} \cite{Zdziarski1996}.
The continuum shape is similar to the soft spectra observed from ULXs \cite{Kaaret2017,Fabrika2021,King2023}.
We find that one gaussian component is capable to describe the line complex around $6.5$~keV (see \ref{fig:nicer_orbit_folded} for more details).
We fit the data with the model 
{\sc constant $\times$ mpl $\times$ (polconst $\times$ reflect $\times$ smedge $\times$ (diskpbb + nthComp) + polconst $\times$ gaussian)} (a similar model was previously used to fit the spectra of SS~433 \cite{Middleton2021} and the incident continuum was found to be similarly soft) and set the parameter rel\_refl$=-1$, which means that we do not take the contribution of the incident X-ray emission into account in the resulting spectra.
For the Main observation the thermal component in the incident spectrum is not needed, so we set its normalisation to zero.
We get an acceptable fit with $\chi^2/{\rm d.o.f.}=936/881$, see Fig.~\ref{fig:specpol_fits} and Table~\ref{tab:specpol}.

The transition from the hard (radio-quiescent) to the intermediate (minor flaring) state in this model is related to the changing shape of the intrinsic continuum, which we nevertheless do not see directly, but only via its reflection.
In order to describe the spectra of the ToO observation, we need a softer incident X-ray spectrum, which is achieved by considering a non-zero contribution of the multicolour disc component.
We get an acceptable fit with $\chi^2/{\rm d.o.f.}=922/885$.
We find that the spectra can also be fit with other models, including those where the primary X-ray emission and reflection both substantially contribute to the X-ray continuum, repeating the diversity of models presented in \cite{Hjalmarsdotter2008}, and confirm that the polarimetric information is vital to choose from a variety of options.
Finally, we note that no currently available public model can account for the complex properties of the reflection in the proposed scenario: we need a convolution model, as we use complex incident spectrum, that considers a hydrogen-poor material and can self-consistently account for the lines.

Spectro-polarimetric modelling for the Main and ToO observations is shown in Fig.~\ref{fig:specpol_fits} and the model parameters are reported in Table~\ref{tab:specpol} (as we are interested in the overall shape of the intrinsic spectra, we note that these parameters are to be considered as guiding, and their set may not be unique). 
In Fig.~\ref{fig:specpol_fits}A we additionally plot the illustrative spectra of (the so-called supersoft and soft) ULXs \cite{Kaaret2017}, which match well the shape of intrinsic spectra of Cyg X-3. The normalisation of intrinsic spectra and ULX spectra are free parameters.
For both the Main and ToO observations, the polarization of the prominent line associated to the complex of neutral iron is found to be consistent with zero, to account for the reduction of measured PD at those energies. 
We also found that a line polarization of $9.4\pm3.0$\% oriented orthogonal to continuum polarization provides an acceptable spectro-polarimetric fit, with $\chi^2=928.13$ for 880 degrees of freedom.
Similarly to an unpolarized line, this can account for the reduction of the observed polarization in the high-energy part of the IXPE energy range.
Albeit the improvement is statistically significant (the F statistic value is 7.2 for a chance probability of 0.7\%), we consider this fit to be affected by the systematic uncertainties of our analysis.
This scenario is also difficult to understand from the physical point of view and therefore we will not discuss it further.
The primary continuum (reflection component) is highly polarized, $22.8\pm 0.4\%$ for the Main observation and $10.0\pm 0.5\%$ for the ToO.

It is well known that Cyg X-3 exhibits a large modulation in flux with the orbital phase of the binary system \cite{vanderKlis1981,Antokhin2022}. 
To investigate possible variations in polarization, we folded the IXPE observations of the source with the ephemeris in Table~2 (2nd model) of \cite{Antokhin2019}.
Phase~0 identifies the superior conjunction of the system, in which the compact object is behind the WR star. 
Data were grouped in 10 (5) phase bins for the Main (ToO) observation and polarization was calculated with the \textsc{ixpeobssim/xpbin} algorithm in three energy bands, 2--3.5, 3.5--6 and 6--8~keV. 
These were chosen to highlight, in the energy range of IXPE, the contributions of the main spectral features identified in the spectro-polarimetric modelling.
The phase-folded PD and PA are shown in \ref{fig:pol_orb_energy_all}, and they display evident orbital variations, which are consistent among the three energy ranges.
PA variations are nearly sinusoidal with an amplitude of $\sim\pm$5\degr, in both the Main and ToO observations, while PD variations are more irregular with an amplitude of a few percent. 
The average PD measured in the ToO is a factor of two lower with respect to the Main observation, and shows similar but not identical orbital profiles.

It is worth noting that, excluding variations due to orbital phase, the polarization remains stable over time.
This is shown in \ref{fig:pol_vs_time}, where we compare the measured polarization degree and angle in the total IXPE energy range, binned with time bins of one period, together with flux variations during the IXPE Main observation. 
While the latter are varying significantly, PD and PA varies around the average value within statistical uncertainties. 
This suggests that the geometry which defines the high polarization observed for Cyg~X-3 is stable with time and essentially unrelated to the ultimate mechanisms producing X-ray variability on superorbital timescales.

\subsection*{Multiwavelength data}\label{sect:meth_mw}

Cyg X-3 has been frequently observed over the past decades from radio through $\gamma$-rays.
On long, weeks to months, time-scales, the source evolves through the sequence of distinct X-ray and radio spectral states (see \ref{fig:diagram} and \cite{Szostek2008_states,Koljonen2010}).
The most frequent state is the hard X-ray, radio quiescent state, which corresponds to the lowest observed X-ray flux.
We observed the source in this state during the Main IXPE run (\ref{fig:broadband_spectra}).
The absorption within the binary is uncertain, hence different branches of spectral models, corresponding to different geometries and dominant spectral components, have been proposed \cite{Hjalmarsdotter2008,Zdziarski2010}, including the models where the incident power-law-like Comptonization spectrum is heavily absorbed or down-scattered in the stellar wind, models with non-thermal Comptonization produced by a steep electron distribution and models with the dominance of the reflection component, in the geometry where the reflector partially covers the primary X-ray source.
The diversity of alternatives prevented firm conclusions on the  intrinsic luminosity in this state, always found to be of the order of $10^{38}$~erg~s$^{-1}$, but precise numbers vary by a factor of 4--5, depending on the model.
At the same time, the uncertainty on the mass of the compact object \cite{Vilhu2009,Zdziarski2013,KoljonenMaccarone2017,Antokhin2022,Suryanarayanan2022}, along with its nature, a neutron star or a black hole, as well as the chemical composition of the hydrogen-poor matter dragged from the WR companion make the estimates of the Eddington luminosity likewise varying.
It has therefore been unclear what kind of accretion regime to expect in this state.

The source occasionally displays spectral transitions to the soft state, followed by the increase of the soft X-ray luminosity and suppression of the radio emission (to the levels below those observed in the quiescent radio state).
Changes of spectral shape have been attributed to the changes of the accretion-ejection geometry.
The X-ray spectra of soft and ultrasoft states resemble thermal emission of the multicolour accretion disc \cite{SS73}, typically seen at luminosities between the Eddington limit and down to $10\%$ of that.
After the transition, a major radio flare may happen, when the highest radio fluxes among all X-ray binaries can be reached \cite{Corbel2012,Egron2021}. 
The second IXPE run was triggered as a target of opportunity observation following the increase of the soft X-ray and drop of the radio fluxes, when the source transited to the suppressed radio state.
IXPE caught the source after the radio recovered, in its intermediate X-ray state, during the minor flaring radio episodes (\ref{fig:diagram}) \cite{Szostek2008_states}.

On shorter timescales, prominent orbital variability of X-ray, $\gamma$-ray, IR and radio fluxes \cite{Parsignault1972,Bonnet-Bidaud1981,Mason1986,Zdziarski2018,Antokhin2022}, as well as X-ray and IR line shapes \cite{Fender1999,Stark2003,Vilhu2009,Kallman2019} have been observed.
Our multiwavelength observations show orbital flux variations in all bands (figs.~\ref{fig:xray_lc} and \ref{fig:radio_lc}).
The lack of apparent variations in the absorbing column density over the orbital phase (noted in \cite{Kallman2019}), suggests that the mean orbital modulation is not primarily caused by the line-of-sight absorption variations of the low-density and partly ionized gas, such as WR wind.
Either scattering in an ionized gas or an asymmetric geometry of a reflector \cite{WhiteHolt1982,Kallman2019} might be more important.
At the same time, the hard-state modulation amplitude has been found to depend on the X-ray energy \cite{Zdziarski2012}, suggesting the scattering alone may not be responsible for the variations.
X-ray orbital profiles are asymmetric \cite{Willingale1985,Antokhin2022}, and the phase of the minimal X-ray flux does not necessarily coincide with superior conjunction (compact object behind the WR star).
Recent study suggest that these phases are close, though, $\phi_{\rm sc}=-0.066\pm0.006$ \cite{Antokhin2022}.
In \ref{fig:nicer_orbit_folded} we show the evolution of the lower-energy spectra observed with NICER throughout the orbital phases during the October-November multiwavelength campaign.  
In broad agreement with previous results\cite{Kallman2019},   we find that changes of spectra as a function of the orbital phase do not follow a simple pattern of changing absorption, as in this case the spectral shape is expected to change substantially. 
Instead, we mostly see variations of the spectral normalisation, which are more in line with changes in the characteristic reflection angle \cite{Poutanen1996}.

At all phases, the energies 6--8~keV are dominated by the complex of the iron emission lines (Fe K lines).
It consists of the neutral iron, iron {\sc xxv} and {\sc xxvi} \cite{Vilhu2009,Kallman2019,Suryanarayanan2022}.
Behaviour of these lines with the orbital phase varies, allowing one to relate the hydrogen-like iron with the compact object \cite{Vilhu2009}.
Analysis of the ratios of the forbidden, resonance and intercombination lines indicates that these species are located in a dense medium, which nevertheless has high ionization \cite{Kallman2019}. 
Interestingly, the Chandra/HETGS spectrum of Cyg X-3 is so far the only fully resolved Fe K complex in an astrophysical source \cite{Suryanarayanan2022}.

\subsubsection*{Supporting X-ray and gamma-ray observations}\label{sect:meth_mw_xrays}

Contemporaneous observations of Cyg X-3 during the Main run have been performed with NICER.
NICER is a soft X-ray instrument onboard the International Space Station (ISS), launched in June 2017.
It consists of 56 co-aligned concentrator X-ray optics, each of which is paired with a single silicon drift detector.  
It is non-imaging, but offers large collecting area, and provides unmatched time resolution in the soft X-ray bandpass,  and sensitive across 0.2--12~keV.
NICER provided monitoring during the IXPE campaign, observing Cyg X-3 between MJD~59884 and 59887.
The resulting average and orbital-phase resolved spectra are shown in \ref{fig:broadband_spectra} and \ref{fig:nicer_orbit_folded}. 
Fluxes were obtained from best fits to these time-sequential data using an analogous model to that from the spectro-polarimetric fits in Table~\ref{tab:specpol}.
NICER has good capabilities for timing studies.
We checked for the presence of short-term (of the order of seconds) variability, but did not find any significant intrinsic fluctuations above the noise level.
This is in line with previous findings \cite{Axelsson2009}.

The Mikhail Pavlinsky ART-XC telescope carried out one observation of Cyg X-3 on 2022 November 4 (MJD~59887) simultaneously with IXPE, with an 86~ks net exposure.
ART-XC is a grazing incidence focusing X-ray telescope on board the Spectrum-Rontgen-Gamma observatory (SRG, \cite{Sunyaev2021}). 
The telescope includes seven independent modules and provides imaging, timing and spectroscopy in the 4--30 keV energy range with a total effective area of $\sim 450$~cm$^2$ at 6~keV, angular resolution of 45 arcsec, energy resolution of 1.4 keV at 6~keV and timing resolution of $23\mu$s \cite{Pavlinsky2021}.
ART-XC data were processed with the analysis software ARTPRODUCTSv1.0 and the CALDB  (calibration data base) version 20220908. 
The ART-XC observation was performed approximately one day before the first NuSTAR observation (Main), therefore spectral parameters measured by ART-XC are close to the ones determined from NuSTAR data with the flux of $\sim4.6\times10^{-9}$ erg~cm$^{-2}$~s$^{-1}$ in the 4--30 keV energy band. 

INTEGRAL observed Cyg X-3 simultaneously with IXPE two times: the first observation lasted from 2022 November 1 21:11 to 2022 November 5 20:23 UT; the second observation lasted from 2022 December 25 02:37 to 2022 December 25 14:53 UT.
Our data analysis is focused on ISGRI, the low energy part of the IBIS telescope \cite{Ubertini99,Lebrun03}.
The INTEGRAL data were reduced using the latest release of the standard On-line Scientific Analysis (OSA, version 11.2), distributed by the INTEGRAL Science Data Centre (ISDC, \cite{Courvoisier03}) through the multi-messenger online data analysis platform (MMODA, \cite{Neronov21}). 
The ISGRI spectra were extracted in the range 30--150 keV with a response matrix with 16 standard channels. 
The spectra of the  first and the second observations were fitted with a simple power law with photon index of 3.6$\pm$0.1 ($\chi^2/{\rm d.o.f.}=18.3/21$) and 3.4$\pm$0.1 ($\chi^2/{\rm d.o.f.}=22.4/21$), respectively.
The fluxes in the range 20--100 keV are 1.7$\times$10$^{-9}$ and 1.4$\times$10$^{-9}$ erg~cm$^{-2}$~s$^{-1}$, respectively.

The Fermi/LAT data on Cyg X-3 was collected during 2022 October 10 to October 21 (MJD~59862--59873) in 0.1--500~GeV energy band. 
Fermi is located at a low-Earth orbit with 90~min period and normally operates in survey mode, which allows the instrument to cover the whole sky in approximately 3~h (see full details of the instrumentation in \cite{Atwood2009}).
The standard binned likelihood analysis~\cite{mattox96} was performed with the latest available \texttt{Fermitools v.2.0.8} software.
The analysis was carried out using the latest Pass 8 reprocessed data (P8R3) \cite{atwood2013pass} for the SOURCE event class (maximum zenith angle $90\degr$) taken at the region centered at Cyg X-3 coordinates. 
The analysis is based on fitting of the spatial/spectral model over a $14\degr$-radius region around the source.
The model of the region included all sources from the 4FGL DR3 catalog~\cite{4fgl_catalogue}, as well as components for isotropic and galactic diffuse emissions given by the standard spatial and spectral templates \texttt{iso\_P8R3\_SOURCE\_V3\_v1.txt} and \texttt{gll\_iem\_v07.fits}.

The spectral template for each 4FGL source present in the model was selected according to the catalog. 
The normalisations of the spectra of all sources, as well as the normalisations of the Galactic diffuse and isotropic backgrounds, were assumed to be free parameters.
We note also that Cyg X-3 is present in the 4FGL catalog as 4FGL J2032.6+4053 a point-like source with a log-parabola-type spectrum.
Following the recommendation of the Fermi-LAT collaboration, we performed the analysis with energy dispersion handling enabled.
To minimize the potential effects from the sources present beyond the considered region of interest, we additionally included into the model all the 4FGL sources up to $10\degr$ beyond this radius, with all the spectral parameters fixed to the catalog values.
The results of the described analysis performed in relatively narrow energy bins are shown in \ref{fig:broadband_spectra}.
The source was not detected in any of the selected energy bins at higher than $2\sigma$ significance (test-statistic 4.0).
The shown upper limits correspond to 95\% false-chance probability and were calculated with the help of \texttt{IntegralUpperLimit} python module, provided within \texttt{Fermitools}.

Cyg X-3 was also observed in $\gamma$-rays with Astrorivelatore Gamma ad Immagini LEggero (AGILE).
AGILE satellite \cite{Tavani2009AGILE} is a space mission of the Italian Space Agency (ASI) devoted to X-ray and $\gamma$-ray astrophysics, operating since 2007 in a low Earth equatorial orbit. 
AGILE in its spinning observation mode performs a monitoring of about 80\% of the entire sky with its imaging detectors every 7 mins.
The data collected with the $\gamma$-ray imager (GRID, 30 MeV--50 GeV), has been analyzed over the periods of MJD~59866--59871, 59883--59889 (Main) and MJD~59938--59942. 
The data analysis was carried out using the last available AGILE-GRID software package (Build 25), \verb+FM3.119+ calibrated filter, \verb+H0025+ response matrices, and consolidated archive (\verb+ASDCSTDk+) from the AGILE Data Center at SSDC \cite{pittori2019}.
We applied South Atlantic Anomaly event cuts and $80^{\circ}$ Earth albedo filtering, by taking into account only incoming gamma-ray events with an off-axis angle lower than $60^{\circ}$.
Flux determination was calculated using the AGILE multi-source likelihood analysis (MSLA) software \cite{bulgarelli2012} based on the Test Statistic (TS) method \cite{mattox96}.
We performed the MSLA for Cyg X-3 by including, as background sources, the 3 nearby pulsars of the Cygnus region (PSR J2021+3651, PSR J2021+4026 and PSR J2032+4127), which are known to be persistent and intense $\gamma$-ray emitters, located at angular distances smaller than $5\degr$ from the source.
For the background sources, we assumed the long-term integration spectra, as reported in the 2AGL Catalog \cite{bulgarelli2019}. 
We modelled the $\gamma$-ray spectrum for Cyg X-3 with a simple power law with a standard 2.0 photon index.
The source was in the quiescent and intermediate state during the time of IXPE observations, hence no prominent $\gamma$-ray activity has been detected. 
The full-band AGILE-GRID upper limits are given in Table~\ref{tab:mw_xray} and are consistent with the Fermi/LAT limits. 
Spectral ULs (50 MeV--3 GeV) are shown in \ref{fig:broadband_spectra}.

\subsubsection*{Supporting radio and submillimeter observations}\label{sect:meth_mw_radio}

Monitoring of Cyg X-3 at radio wavelengths contemporaneous with IXPE was performed using the Large Array of the Arcminute MicroKelvin Imager (AMI-LA), RATAN-600, Medicina, Effelsberg, upgraded Giant Metrewave Radio Telescope (uGMRT) and Submillimeter Array (SMA) telescopes.
This coverage allowed us to identify the state of the source, produce the broadband spectrum and make constraints on the PA at longer wavelengths.
A summary of these observations can be found in Table~\ref{tab:mw_radio} and in \ref{fig:radio_lc}.

Cyg X-3 was observed at 15.5~GHz with the AMI-LA \cite{Zwart2008,Hickish2018} during the IXPE observing campaigns.
The AMI-LA consists of eight 13-m antennas, which measure one polarization (Stokes $I+Q$), over a wide bandwidth of 12 to 18~GHz in 8 broad channels. 
The observations were usually $\sim 1$-hr long, with some longer observations, up to $\sim 6$~hr, from Nov 3rd to 6th. 
Each observation consisted of 10-min scans on Cyg X-3, interleaved with short observations of a nearby compact calibrator source J2052$+$3635, which was used to apply phase corrections, and monitor the sensitivity of the telescope. 
The data were processed using standard procedures: (i) to automatically eliminate bad data due to various technical problems and interference; (ii) manually edit any remaining interference (which included the end channels, which were more prone to interference), and periods of heavy rain; (iii) use the interleaved observations of J2052$+$3635 to provide the initial phase calibration of each antenna in the array throughout each observation, (iv) set the overall flux density scale by comparison with daily observations of the standard calibrator source 3C~286, together with the ``rain gauge'' measurements made during the observations to correct for varying atmospheric conditions \cite{Zwart2008}. 
Flux densities at 15.5~GHz were derived for 10-min averages, from the central 6 broad frequency channels (i.e.\ covering $13.6{-}17.4$~GHz).
The resulting light-curves are shown in \ref{fig:radio_lc}A and D.

To monitor Cyg X-3 we triggered a Target-of-Opportunity program with the 32-m Medicina radio telescope in order to follow the evolution of the radio emission during the IXPE observations.
We carried out observations at the central frequency of 8.4 GHz (X-band) with the Total Power continuum backend on 2022 October 14--18. 
Each session lasted 5~h per day in order to track the fast flux density variations even during the quiescent state. 
We performed On-The-Fly cross-scans and maps along the Right Ascension and Declination directions, setting a bandwidth of 230~MHz to avoid the strongest radio frequency interference (RFI). 
Scans were performed along a length of $0.7\degr$ at a velocity of $2.4\degr/$min at 8.4 GHz, with a sampling time of 40\,ms. Data were calibrated through repeated cross-scans centered on NGC 7027 at different elevations.
This calibrator has the advantage to be very close in elevation to the target.
We extrapolated the calibrator flux density according to \cite{Ott1994}.
The calibration procedure included the corrections for the frequency-dependent gain curves, in addition to the compensations for the pointing offset measured on each scan. 
The data analysis was performed with the Single-Dish-Imager, a software designed to perform automated baseline subtraction, radio interference rejection and calibration \cite{Egron2017}. 
We estimate the final accuracy of our measurements to be $\sim 8\%$ at 8.4 GHz.
The resulting light curve is presented in \ref{fig:radio_lc}B.

Observations of Cyg X-3 were also performed with the 100-m Effelsberg dish on 2022 October 9, 13, 14, and 18 with the S45mm-receiver and the spectropolarimeter backend.
Acquisitions were performed over two bands, 5.4--7.2 ($f_{\rm center}=6.3$~GHz) and in two subbands of the second band 7.6--8.2 \& 8.4--9.0~GHz ($f_{\rm center}=8.3$~GHz).
These frequency ranges (especially the omission of the center part of the second band) were chosen to avoid RFI.
We measured the flux density with the cross-scans-method, doing several subscans in azimuth and elevation (12 in the case of Cyg X-3).
All subscans were corrected for pointing offsets and averaged.\
After that the atmospheric absorption and the loss of sensitivity due to gravitational deformation of the dish were corrected (both effects are rather small).
The final calibration was done via suitable flux density calibrators (i.e. 3C~286 and NGC~7027). 
For the polarization, instrumental effects were corrected by a M\"{u}ller matrix method.
A number of calibrators were observed before and after the actual observations of Cyg~X-3, to determine the various effects properly.
No polarization was detected in the Effelsberg data meaning that the level of polarization must be lower than 5\%.
The resulting light-curves are shown in fig~\ref{fig:radio_lc}B and C.

Cyg X-3 was monitored at 4.7 and 8.2 GHz on a daily basis \cite{Trushkin2023} at the North sector of the RATAN-600 telescope using the uncooled tuned receiver in the total power radiometer mode \cite{Tsybulev2018}.
This mode allows one to perform sensitive observations, with precision being limited by the presence of RFI.
Typical accuracy of 5\% for fluxes near 100 mJy has been reached during the contemporaneous observations with IXPE.
The main parameters of the antenna (effective area and beam size) were calibrated with the source NGC~7027.
Observations of NGC~7027 in the multi-azimuthal mode gave a flux density of 5.38 Jy at 4.7 GHz, in agreement with the standards \cite{Ott1994}.
Additional intraday observations of Cyg X-3 at 4.7 and 8.2~GHz were carried out with the ``Southern sector and Flat mirror'' configuration.
The increased field of view ($\pm30\degr$, as compared to the observations with the North Sector) in this configuration allowed one to follow the source longer. 
For discrete antenna configurations (with step $2\degr$) we carried out 31 measurements, taken every 10 minutes.
The resulting light-curves at 4.7 and 8.2 GHz are shown in \ref{fig:radio_lc}B and C for the Main run, and in \ref{fig:radio_lc}E for the ToO run.
 
Observations of Cyg X-3 during the Main IXPE observation with uGMRT were performed following the Director Discretionary Time (DDT) requested.
Due to scheduling constraints, the observations were only granted on November 2 and 3  for $\sim 5$~h each, i.e. a full orbit.
Observations were performed at Band 5 (1--1.4 GHz) using a correlation bandwidth of 400 MHz and 2048 frequency channels.
The observing strategy featured cross-scans on the source interleaved with calibrators for phasing and flux references.
The absolute flux density scale is tied to the Perley-Butler 2017 scale.
The CAPTURE pipeline \cite{kale21} was used to analyse the GMRT data.
The error on the total flux density of the source includes the error on the Gaussian fit and an absolute flux density error of 10\% added in quadrature.

 Cygnus X-3 was observed by the SMA located on Maunakea  in Hawaii on  2022 October 19 and 2022 November 2. 
The SMA observations use two orthogonally polarized receivers, tuned to the same frequency range in the full polarization mode. These receivers are inherently linearly polarized but are converted to circular using the quarter-wave plates of the SMA polarimeter \cite{Marrone2008}.
The lower sideband (LSB) and upper sideband (USB) covered 209--221 and 229--241 GHz, respectively. 
Each sideband was divided into six chunks, with a bandwidth of 2~GHz, and a fixed channel width of 140 kHz.
The SMA data were calibrated with the MIR software package.
Instrumental polarization was calibrated independently for the USB and LSB and removed from the data.
The polarized intensity, PA and PD were derived from the Stokes $I$, $Q$, and $U$ parameters. 
MWC 349 A and BL Lac were used for both flux and polarization calibration and Neptune was used for flux calibration.
Observations on October 19  were done with four antennas and a median 225 GHz opacity of $\sim$0.2 while those on 2 November were obtained with seven antennas and a median opacity of $\sim$0.1. 
Due to the low level of polarization the overall polarization measurements are of low statistical significance, especially for the October 19  observation.
For the October 19  observation it was necessary to exclude 1 of the four antenna and in November 2nd observation data after UT 9.2 was excluded due to significant increase in phase instability as a result of weather conditions.
The overall flux uncertainty in an absolute sense is $\sim$5\% of the continuum flux value.
The values shown in Table~\ref{tab:mw_radio} are averages over the entire observation of that day.
Light curves of the total intensity (Stokes $I$) for the two days are shown in fig~\ref{fig:radio_lc}A.

\subsection*{Modelling}\label{sect:meth_model}

\subsubsection*{Analytical modelling of the funnel}\label{sect:meth_model_analyt}

At high accretion rates, the accretion disc possesses a critical point, the spherization radius, at which the matter can leave the disc pushed by radiation pressure forces \cite{SS73,Poutanen07}.
It forms an axially symmetric outflow with an empty funnel around the disc axis. 
Radiation emitted by the accretion disc cannot escape freely, but is collimated by the funnel walls.
As a result an observer looking along the funnel will see strongly amplified emission \cite{King2001}.
On the other hand, an observer located at a large inclination angle to the axis sees the photosphere that is situated at a significant distance from the central source, which depends on the mass loss rate that in turn depends on the accretion rate \cite{Poutanen07}.
Such an observer can see radiation scattered and reflected from the funnel walls at high elevations, where the matter is mostly neutral (flux from the central source is expected to be reduced due to small solid angles of matter occupied on the plane of the sky and because of the collimation of the incident emission at the bottom of the funnel).

We approximate the funnel geometry by the truncated cone (see Fig.~\ref{fig:model_funnel}A and \ref{fig:model_alpha_rho}A), which has two main parameters: $R$, the distance to the X-ray photosphere where the optical depth becomes comparable to unity, scaled to the inner radius of the outflow in the accretion disc plane, and the angle $\alpha$ at which the upper boundary of the funnel is seen from the primary X-ray source.
Unpolarized radiation emitted by the central source (which is the inner accretion disc and the collimated radiation from the inner part of the funnel) is impinging on the  wall higher up in the funnel. 
The probability for photons to be reflected is proportional to the energy-dependent single-scattering albedo $\lambda_E$, which is the ratio of the scattering opacity to the total (scattering and photo-electric) opacity. 
Because in the IXPE range $\lambda_E\ll 1$, the reflected radiation is dominated by single-scattered photons (multiple scattering orders contribute to the observed spectrum if $\lambda_E\sim1$).
This radiation is polarized with the PD for Thomson scattering (valid in the IXPE range) being dependent on the cosine of the scattering angle $\mu$ as  
\begin{equation}\label{eq:pd_thomson}
P(\mu) =  \frac{1-\mu^2}{1+\mu^2} . 
\end{equation} 
The PA of this radiation, which we denote as $\chi_0$, lies perpendicular to the scattering plane. 
The intensity of reflected radiation is proportional to the phase function $\frac{3}{4}(1+\mu^2)$ and the ratio $\eta_0/(\eta+\eta_0)$ (page 146 in \cite{Cha60}), where $\eta_0$ is the cosine of the angle between the local normal to the funnel wall and the incoming radiation beam, while $\eta$ is the cosine of the angle between direction to the observer and the normal. 
Thus the Stokes parameters representing  the linearly polarized reflected radiation are 
\begin{equation} 
\left( \begin{array}{c} 
I_E \\  Q_E  \\ U_E   
\end{array} \right) = 
\lambda_E
\frac{3}{4}(1+\mu^2) 
\frac{L_E}{4\pi r^2} \ \left( \begin{array}{c} 
1 \\  P \cos2\chi_0 \\  P \sin2\chi_0   \end{array}  \right)  \frac{\eta_0}{\eta+\eta_0},
\end{equation}
where $L_E$ is the luminosity of the central object and $r$ is the distance from the center to the element of the funnel. 
Integrating this expression over the visible surface of the funnel, we get the observed flux and the corresponding Stokes parameters.   
We see that all Stokes parameters in the single-scattering approximation are proportional to $\lambda_E$ and therefore the PD of the total radiation is energy-independent. 

A natural condition for the primary source to be obscured is $i>\alpha$.
In Fig.~\ref{fig:model_funnel}B we show the contours of constant PD as a functions of $\alpha$ and $i$, for a chosen $R=10$.
Two branches of solutions are possible for $i\lesssim40\degr$: the lower branch with a narrow funnel $\alpha\sim10\degr$, and an upper branch with $\alpha\approx i$, where the observer looks almost along the funnel walls.
We note tightly-packed contours near this branch, indicating that any small, a few degrees, variation of the opening angle would cause changes in the observed PD by tens of per cent.
In contrast, the time dependence of the observed PD, averaged over orbital phase, is consistent with being constant, with the standard deviation of $2.5\%$ (see \ref{fig:pol_vs_time}).

In \ref{fig:model_alpha_rho}B we show the dependence of PD parameters $\alpha$ and $R$, for $i=30\degr$ (for other possible values of $i$, the topology and resulting numbers are similar).
We see the same two branches of a possible solution corresponding to PD=23\% and consider only the lower one for the aforementioned reason (small observed variations of PD).  
The part of the diagram below $R=1/\sin\alpha$ is forbidden, because it corresponds to $\rho<1$, i.e. a converging towards the axis outflow. 
For the observed PD, the minimum possible size of the photosphere is $R=8$, which corresponds to $\alpha=8\degr$. 
At a larger $R$, the required $\alpha$ increases, saturating at $\approx15\degr$.  

The computed PDs in \ref{fig:model_funnel}B and \ref{fig:model_alpha_rho}B correspond to the case when polarization is produced solely at the inner surface of the funnel, which can be realised for a very high Thomson optical depth.
These conditions can be applicable to the Main observation.
Changes of polarization properties in the ToO observation can be caused by the reduction of the Thomson optical depth of the funnel.
In this case, we expect to see scattered radiation from some volume around the funnel walls, rather than solely from its inner surface.
This leads to the increased role of photons scattered at small angles, hence reduction of the net polarization.
Alternatively, the scattering may proceed right above the funnel, in the optically thin WR wind.
Our estimates of the Thomson optical depth from the mass loss rate and wind velocity, assuming hydrogen-poor material \cite{Antokhin2022} give $\tau_{\rm T, WR}\sim0.1-0.5$.
For small optical depth, $\tau_{\rm T, WR}\approx0.1$, the spectrum of the scattered radiation closely resembles that of the incident continuum. 
However, the observed spectral shapes do not correspond to the spectra of any other unobscured accreting source. 
For larger $\tau_{\rm T, WR}\sim 1$, on the other hand, the effect of multiple scattering tends to decrease the PD towards higher energies within the IXPE range, which is not consistent with the results of our spectro-polarimetric modelling. 
Thus, to be consistent with the data, this scenario requires tight constraints $\tau_{\rm T, WR}\approx 0.3-0.5$, which might be hard to realise. 
Our calculations show that the resulting PD in this scenario is nearly independent of the funnel angle $\alpha$ at any inclination $i>\alpha$, hence this case cannot account for the change of PD between the Main and ToO observations, as changes of the $\tau_{\rm T, WR}$ within the allowed narrow range would only lead to variations of flux, but not the PD.
If we consider this scenario for the ToO observation, then the observed PD$\approx10\%$ translates to the inclination $i\approx25\degr$ according to equation~(\ref{eq:pd_thomson}).

An important property of the observed X-ray polarization is its prominent orbital phase-dependent variation (Fig.~\ref{fig:qu_orb}).
Interestingly, the polarization is mostly ``misaligned'' from the East-West direction (i.e., from the orbital plane) during the phases of inferior and superior conjunctions, when the left-right directions (that give non-zero contributions to the Stokes $U$) are expected to be symmetric in the simple picture with the cone-shaped funnel pointing in the direction of the orbital axis (see orbital flux and PA profiles in  \ref{fig:pol_orb_energy_all}D,E,F). 
In the proposed scenario, the outflow from the compact object is expected to collide with the wind of the WR star, resulting in an asymmetry of the funnel and its surroundings.

We first considered geometries where the funnel is shaped as an oblique, truncated cone and also modelled a situation where the funnel axis is not aligned with the orbital axis.
In both cases, the orbital variations arise from the asymmetry of the funnel itself.
The first model does not reproduce the strength of the signal in Stokes $U$ because for a narrow funnel, most of the reflected photons that reach the observer are scattered at nearly the same angle, even for the additional part of the funnel surface producing geometrical asymmetry.
The second case is reminiscent of the rotating vector model, which has a tight relation between the PD and PA variations \cite{Radhakrishnan1969,Poutanen2020}.
In order to reproduce the phase shift between the observed PD and PA variations, we find that the funnel should be inclined in the direction of movement in the orbit.
On the contrary, the funnel moving through the stellar wind is expected to be tilted in the direction opposite to its velocity vector (at least, its optically thin outer parts, which effectively contribute to the observed emission and polarization).
Hence, we conclude that the variations in $U$ are not caused by the asymmetric shape of the funnel itself.

We also considered an alternative scenario where the orbital modulation is produced via scattering of the primary X-ray emission in the point where the matter from the companion hits the accretion disc, the scenario proposed for the orbital variability of Cyg X-3 system in analogy to the low-mass X-ray binary systems at high inclinations \cite{WhiteHolt1982}.
The impact point contains a heated material, which can rise above the orbital plane, producing a regular attenuation/dipping pattern.
The solid angle of this material can be expected to be comparable to the size of the companion star, and the variations of produced by this component are smaller that $\sim1\%$ (e.g., modelling and discussion in \cite{Rankin2024}).
In contrast, an asymmetric structure occupying substantial solid angle, as seen from the X-ray source, is required in order to reproduce the observed variations of polarization.

The accretion geometry of Cyg X-3 and other high-mass X-ray binaries contains a common component, the bow shock produced by the movement of the compact object through the wind of the companion.
The outflow from the compact object is expected to collide with the wind of the WR star, producing an enhanced density region.
The presence of the bow shock in Cyg X-3 has been exploited to explain the orbital changes of X-ray and IR fluxes \cite{Antokhin2022}.
We test whether this component can be responsible for the prominent polarimetric variations.
In contrast to the beamed X-ray emission escaping along the funnel, the reflected and reprocessed light of the funnel walls is more isotropic.
We suggest that the high-amplitude orbital variability of PA seen both during the Main and ToO observations is produced thanks to the scattering of the scattered and reprocessed radiation of the funnel walls from the inner surface of the bow shock.
A fraction $\eta_{\rm bow}$ of the funnel radiation is scattered by the bow shock.
We approximate its surface by a cylindrical sector parameterised by the angular extent $\phi_{\rm cyl}$, the azimuth of its center at phase 0 (superior conjunction) relative to the line connecting the stars $\phi_{\rm cen}$, and by the height-to-radius ratio of the cylinder $H_{\rm cyl}/\rho_{\rm cyl}$. 
In this combined geometry with the funnel and the bow shock, the average polarization comes from the radiation reflected from the funnel (described by the parameters $\alpha$ and $R$) and the orbital variability arises from the scattering of the mostly isotropic radiation off the inner surface of the bow shock (with parameters $\eta_{\rm bow}$, $\phi_{\rm cyl}$ and $H_{\rm cyl}/\rho_{\rm cyl}$).
We note that the same geometry description can also be applied to the scenario where orbital variations are caused by scattering from the accretion stream/bulge.
In this case, the scattering material should have a small azimuthal extent, $\phi_{\rm cyl}<90\degr$.

In \ref{fig:bow_shock} we show an example of description of orbital variations for parameters $\alpha=10\degr$, $R=50$, $H_{\rm cyl}/\rho_{\rm cyl}=1$, $\phi_{\rm cyl}=220\degr$, $\phi_{\rm cen}=90\degr$ (at superior conjunction, the center is located to the left of the line connecting the stars) and $\eta_{\rm bow}=0.09$. 
We see, however, that the model does not reproduce the shape of the PD exactly, and attribute this to the simplicity of the assumed bow shock geometry.
For our parameter $\phi_{\rm cen}=90\degr$, the bow shock is located at maximal angular distances from the plane formed by the observer, the WR star and the compact object at superior and inferior conjunction.
In other words, we expect the PA to be maximal/minimal at the conjunctions and cross its average value, $\sim90\degr$, close to quadratures.
From the fact that PA is maximal in the first orbital bin, we deduce that the Cyg X-3 system rotates in the counterclockwise direction.

\subsubsection*{Monte-Carlo modelling of the toroidal envelope}\label{sect:meth_model_mc}

To consider the effects of finite optical depth and dependence of the resulting polarization on geometry, we ran Monte-Carlo (MC) simulations using the code {\tt STOKES} version 2.07 \cite{Goosmann2007,Marin2018b}.
The code traces polarization of photons propagating in media, taking into account the effects of photoelectric absorption and Compton down-scattering.
Both continuum and line emission are considered.
Our goal is to identify the parameter space for which the average observed polarization can be reproduced.
We refer to \cite{Podgorny2023sub} where details of the modelling are given and where the results for broader parametric range and more distinct geometry cases are shown.
We choose to show the reprocessing in an elliptical torus (see Fig.~\ref{fig:MC_geometries}A for a geometry sketch), which represents an alternative geometry from the cone-shaped outflow that we described in the previous section (even though this geometry might not be directly applicable to the super-Eddington outflow configuration).
The profile of the elliptical torus is parameterised through the cylindrical distance $\rho$, the grazing angle $\alpha$ that is similar to the opening angle of the funnel in the cone geometry, and the minor axis $b$.
Only the ratios of distances affect polarization properties.
The point-like source located at the center of the coordinate system illuminates the axially symmetric scattering region.

The densities and atomic properties within the equatorial scattering region are homogeneous; the column density along the scatterer is proportional to the length of the scattering region between the source and the observer.
We assume solar abundance from \cite{Asplund2005} with $A_\mathrm{Fe} = 1.0$, but we remove the neutral hydrogen from the absorbers due to the expected hydrogen-poor environment \cite{Fender1999, Kallman2019}.
The main parameters of the medium that control polarization properties are thus neutral helium number density, being the most abundant element, which is expressed through the column density $N_{\rm He}$ between the center and an equatorial observer, and the number density of free electrons in the medium (related to ionization), defined through the equatorial electron-scattering Thomson optical depth $\tau_{\rm e}$.
We show the results for the unpolarized primary radiation, but tested various cases of polarized primary emission \cite{Podgorny2023sub}.
The same holds for the primary spectral distribution, which we fix as a power law with the photon index $\Gamma=3$ for simplicity.

As an example, in \ref{fig:MC_geometries}B we show the 2--8~keV integrated PD as a function of the observer inclination $i$ and the ellipse grazing angle $\alpha$ for the case $N_{\rm He}=8.5 \times 10^{23}$~cm$^{-2}$ and $\tau_{\rm e}=7$, corresponding to the partially ionized case with nearly equal number densities of helium and free electrons.
The white dashed curves represent contours corresponding to the PD$\approx 21\pm3$\% in the 2--8~keV range, where the lower and upper limits correspond to the characteristic uncertainties of the simulations.
Cells with PD that fall in the correct range are also highlighted with black rectangles.
We find that the contours form the same topology in the $(i, \alpha)$ space as for the analytical model and give similar, within uncertainties, allowed combinations of $(i, \alpha)$.
We explored the parameter space with various aspect ratios, and compared the multiple-scattering to single-scattering cases.
In all cases, we have been able to obtain a general pattern of two solutions, similar to the two branches in \ref{fig:model_funnel}B.
At 3.5--6~keV we obtained almost no difference between the single-scattering and multiple-scattering cases, as at these energies the single-scattering albedo is low. 
For all values of $\alpha$ and $i$ on \ref{fig:MC_geometries}B, the obtained PA is perpendicular to the axis of symmetry.

The energy dependence of polarization in the 2--8 keV range matches that of the Main observation if spectral lines are included \cite{Podgorny2023sub}.
If the density or ionization of the medium is lower, we obtain the same PA and lower PD, down to an almost unpolarized emission, depending on the modelling subtleties. 
For an order-of-magnitude lower density the obtained results are in line with the data in the ToO observation, supporting the interpretation of the decreased mass accretion (and outflow) rate for this state.

\subsection*{Intrinsic and apparent luminosity estimates}
\label{sec:lumin}

Using the analytical model described above, that was applied to the Main observation, we can compute the luminosity escaping in the direction along the funnel axis $L_{\rm ULX}$ from the observed flux.
We assume that the primary X-ray source within the funnel is isotropic and produces luminosity $L_{\rm X}$.
In this case, the luminosity escaping in a given solid angle is proportional to this solid angle, $L_{\Omega}\propto\Omega$.
Three distinct sites contribute to the total X-ray luminosity: the funnel opening, where the fraction proportional to the solid angle of the funnel escapes, the reprocessing site seen to the observer (region between point $L$ and the upper boundary of the funnel in Fig.~\ref{fig:model_alpha_rho}A) and the lower layers of the funnel (between point $L$ and the disc plane).
The contribution of the latter luminosity may come in the form of a soft reprocessed X-ray radiation and is not clearly visible in our data.
The contribution of the former two can be related to the intrinsic X-ray luminosity:
\begin{equation}
    L_{\rm ULX}  = 
    \frac{2\pi}{\Omega_{\rm ULX}} L_{\rm X},
\end{equation}
where $\Omega_{\rm ULX}=2\pi(1-\cos\alpha)$ is the solid angle of the funnel opening as seen from the primary X-ray source. 
The observer receives the flux $F_{\rm obs}$, which is emitted by (reflected from) the visible part of the inner surface of the funnel (geometry in \ref{fig:model_alpha_rho}A).
The luminosity intercepted by this part can be expressed as
\begin{equation}
    L_{\rm refl} = \frac{\Omega_{\rm refl} }{\Omega_{\rm ULX}} aL_{\rm X},
\end{equation}
where $a$ is the scattering albedo and $\Omega_{\rm refl}$ is the characteristic solid angle of the reflecting surface (that the observer is able to see), as viewed from the primary X-ray source.
The reflected luminosity produces the observed flux we detect, hence $F_{\rm obs}=L_{\rm refl}/(4\pi D^2)$.
Combining the terms, we can get the expression for the luminosity escaping along the funnel:
\begin{equation}\label{eq:l_ulx}
    L_{\rm ULX} = \frac{2\pi}{\Omega_{\rm refl}}\frac{4\pi D^2 F_{\rm obs}}{a}.
\end{equation}
The solid angle of the reflecting surface can be expressed as 
\begin{equation}\label{eq:omega_refl}
    \frac{\Omega_{\rm refl}}{2\pi} = \cos{\alpha} - \cos{\alpha^*},
\end{equation}
where $\alpha^*$ corresponds to the angle at which the lowest interior part of the funnel is seen by an observer (\ref{fig:model_alpha_rho}).
This angle is related to the funnel opening angle $\zeta$ as
\begin{equation}\label{eq:alpha_star}
    \tan{\alpha^*} = \frac{\tan{\zeta}}{1-1/\rho_{\rm L}},
\end{equation}
$\rho_{\rm L}$ is the radius of the funnel at point $L$, in units of inner radii of the outflow.
It can be expressed through the model parameter $\alpha$, the cylindrical radius of the funnel outer boundary $\rho=R\sin{\alpha}$ and the observer inclination $i$ as
\begin{equation}\label{eq:rho_l}
    \frac{1}{\rho_{\rm L}} = \frac{1}{\rho} 
     \frac{\tan{i}+\tan{\zeta}}{\tan{i}-\tan{\zeta}} . 
\end{equation}
The opening angle is in turn related to the model parameter $\alpha$ as
\begin{equation}\label{eq:zeta}
    \tan{\zeta} = (1-1/\rho) \tan{\alpha}.
\end{equation}
Substituting equation~(\ref{eq:rho_l}) and (\ref{eq:zeta}) into equation~(\ref{eq:alpha_star}) and obtaining $\cos{\alpha^*}$, we find $\Omega_{\rm refl}/2\pi$ as a function of the parameters $\rho$, $\alpha$ and $i$.
Further, for the given observed polarization we can relate $\alpha$ and $R$ (see red contour in \ref{fig:model_alpha_rho}B), which makes $\Omega_{\rm refl}/2\pi$ only a function of $\alpha$.
In \ref{fig:model_alpha_rho}C (solid red line) we show that, for all combinations $(\alpha,R)$ which give the observed polarization, we find $\Omega_{\rm refl}/2\pi < 3\times10^{-2}$.

We take the observed flux (that is seen by IXPE, without correcting for the absorption in the WR wind and along the line of sight in the Galaxy) as a lower limit on $F_{\rm obs}=1.5\times10^{-9}$~erg~cm$^{-2}$~s$^{-1}$ (which corresponds to the observed luminosity $L_{\rm obs}=1.6\times10^{37}$~erg~s$^{-1}$, assuming the distance 9.7~kpc).
Albedo is a function of energy, abundance and viewing angle \cite{MagdziarzZdziarski1995}.
Motivated by our spectral fitting  (comparing the intrinsic and reflected 2--8~keV spectra in \ref{fig:specpol_fits}) we take the value $a\sim0.1$ as a conservative approximation.
Inserting the numbers into equation~(\ref{eq:l_ulx}), we get a lower limit on the luminosity seen along the funnel in the 2$-$8~keV range, $L_{\rm ULX}=5.5\times10^{39}$~erg~s$^{-1}$.

To estimate the intrinsic X-ray luminosity, we need to take into account several additional factors.
First, the observed fluxes have to be corrected for absorption.
For our spectropolarimetric modelling, however, we ignored absorption, hence we take $F_{\rm unabs}=F_{\rm obs}$.
Further, we need to take into account the bolometric luminosity correction, $f_{\rm bol}$.
This estimate consists of two contributions: first, the observed NICER (0.5--12~keV) flux is 1.7 times higher than the flux in the IXPE band (Table~\ref{tab:mw_xray}).
Second, the higher-energy, 12--60~keV, part of the incident continuum contains luminosity that is $0.4$ times that in the IXPE range, as computed from the spectral shape found in spectro-polarimetric fitting, Table~\ref{tab:specpol}.
Hence the lower-limit estimate on the flux in the 0.5--60~keV band is $2.1F_{\rm obs}$.
The lower-energy part of the spectrum would contribute more if the spectral peak is achieved at energies below IXPE band (that is indeed expected for the soft intrinsic spectrum) hence we make a conservative estimate on $f_{\rm bol}\gtrsim2-3$.
Finally, we consider the range of luminosities for different pairs of ($\alpha$, $R$) satisfying the observed PD (\ref{fig:model_alpha_rho}).
The intrinsic bolometric X-ray luminosity can be expressed through the unabsorbed X-ray flux as
\begin{equation}
    L_{\rm X, bol} = \frac{4\pi D^2 f_{\rm bol}F_{\rm unabs}}{a}\left( 1 + \frac{\Omega_{\rm ULX}}{\Omega_{\rm refl}}\right) > 3 \times10^{38} \left( 1 + \frac{\Omega_{\rm ULX}}{\Omega_{\rm refl}}\right)
    {\rm erg~s^{-1}}.
\end{equation} 
In \ref{fig:model_alpha_rho}C (blue dashed line) we show the dependence of the amplification factor \mbox{$(1 + \Omega_{\rm ULX}/\Omega_{\rm refl})$} on the angle $\alpha$.
The obtained luminosity can be compared to the Eddington accretion rate for He (given that the source shows hydrogen-poor properties \cite{Fender1999}) $L_{\rm Edd, He}=2.6\times10^{38}(M_{\rm X}/M_{\odot})$~erg~s$^{-1}$ (where $M_{\rm X}$ is the mass of the compact object and $M_{\odot}$ is the solar mass).
For small funnel angles, $\alpha=8\degr$, the intrinsic bolometric X-ray luminosity exceeds the Eddington limit for a compact object with low mass, $M_{\rm X}/M_{\odot}\lesssim2$, such as a neutron star.
Evolutionary arguments however suggest that the Wolf-Rayet-fed compact object should swiftly become a black hole \cite{Lipunov1982,Bogomazov2014}.
For a compact object $M_{\rm X}/M_{\odot}=10$ we find that the bolometric luminosity estimate exceeds the Eddington limit (in He-rich material) for a combination of $\alpha=15\degr$ and $f_{\rm bol}=3$ or $\alpha=16\degr$ and $f_{\rm bol}=2$. 
For values $\alpha>16\degr$ and $f_{\rm bol}>3$ the observed limit exceeds the Eddington limit for $M_{\rm X}/M_{\odot}=20$, which corresponds to the heaviest Galactic black hole mass measured today \cite{Miller-Jones2021}.
Interestingly, for the case when scattering proceeds in the optically thin wind above the funnel, the factor in brackets should be replaced with $1/\tau_{\rm T, WR}\approx2-10$, which does not affect the final estimate of the intrinsic luminosity.

If the source is surrounded by the envelope with a narrow funnel, the primary luminosity will be beamed in the direction along its axis \cite{King2001}.
From the obtained constraints on the funnel half-opening angle $\zeta$ ($\lesssim\alpha$), we can directly get the geometrical amplification (beaming) factor, 
\begin{equation}\label{eq:beaming_f}
  b = \frac{1}{1-\cos\zeta} \gtrsim 30. 
\end{equation} 
The beaming is expected to vary with the opening angle and can ultimately depend on the mass accretion rate \cite{King2009}.
More precise estimates of the beaming factor can be obtained by taking into account the interactions of photons with the funnel walls \cite{Dauser2017,Mushtukov2023}.
Monte-Carlo simulations of multiple reflection and reprocessing events within the funnel show that a substantial fraction of photons leave the system outside of the solid angle $\Omega_{\rm ULX}$, leading to a reduction of beaming factor.
The magnitude of the reduction, in turn, depends on the height of the funnel: larger number of photons leave the system outside of $\Omega_{\rm ULX}$ for larger $R$. 
The estimate in equation~(\ref{eq:beaming_f}) corresponds to the limiting case of infinitely large $R$. 


\newpage 

\section*{Extended Data Figures}

\renewcommand\thefigure{Extended Data Fig. \arabic{figure}}  
\renewcommand{\figurename}{}
\setcounter{figure}{0}

\begin{figure}[h]%
\centering
\includegraphics[width=\textwidth]{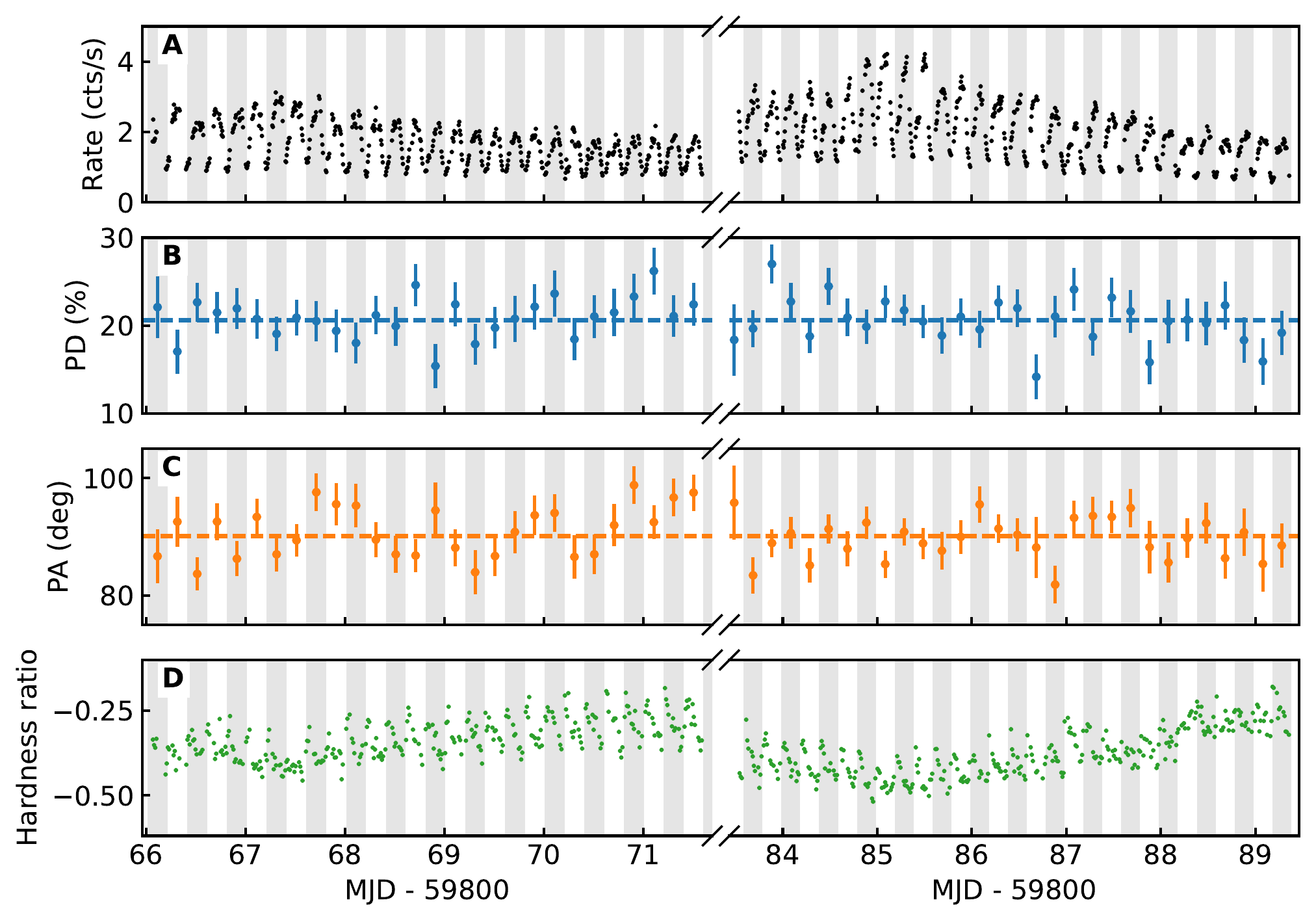}
\caption{\textbf{Variation with time of flux and polarization for the IXPE Main observation. }
(A) The total rate in the 2--8~keV energy range, binned in time intervals of 500~s.
The PD (B) and the PA (C) are averaged over one orbit, as defined by the ephemeris of \cite{Antokhin2019}. Dashed horizontal lines are the average values. 
(D) The hardness ratio defined as the ratio of the difference in the IXPE count rates in the 4--8 and 2--4 keV energy bands to their sum in 1000~s time bins. 
Alternating vertical bands identify different orbits.
Data are presented as mean values over the time bin and the error bars correspond to 1$\sigma$ confidence levels.
} \label{fig:pol_vs_time}
\end{figure}

\begin{figure}
\centering
\includegraphics[width=0.7\textwidth]{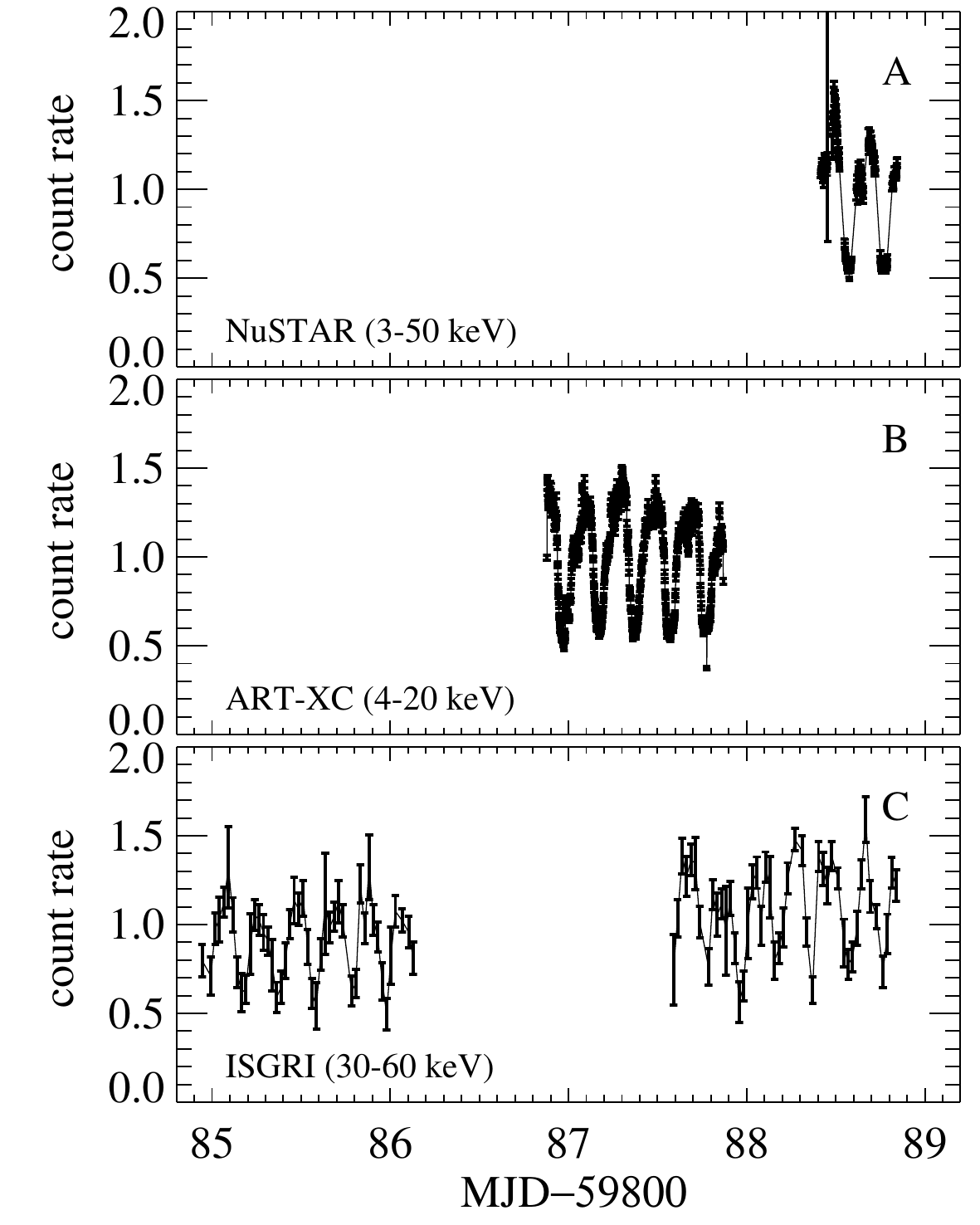} 
\caption{\textbf{X-ray light curves of Cyg X-3.} 
X-ray count rates normalised to the average during the Main observation obtained by three X-ray telescopes: NuSTAR (A), SRG/ART-XC (B) and INTEGRAL/ISGRI (C).
The IXPE exposure covers the entire duration of the displayed observations.
}\label{fig:xray_lc}
\end{figure}

\begin{figure} 
\centering
\includegraphics[width=0.9\textwidth]{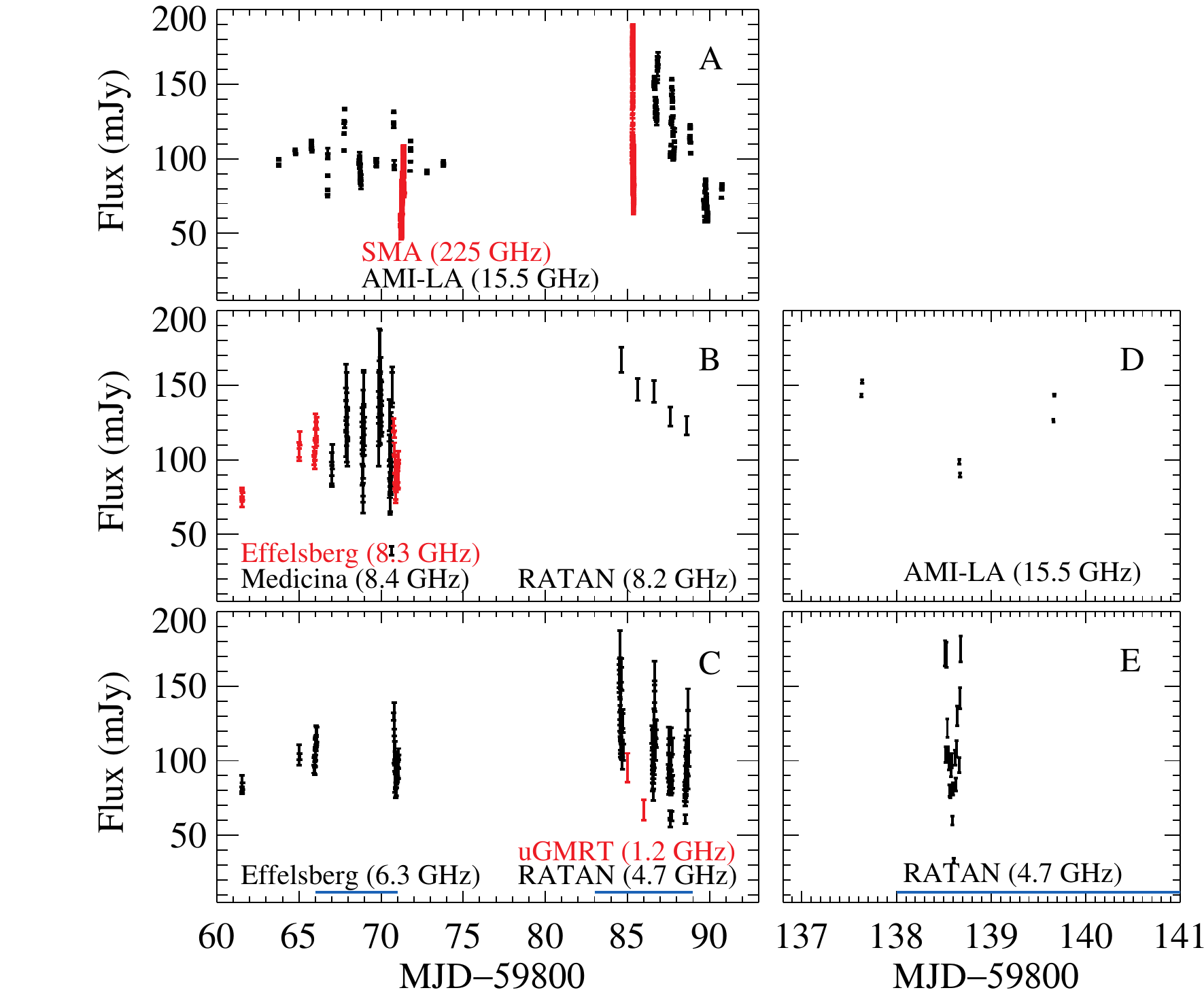}
\caption{\textbf{Radio and sub-mm light curves of Cyg X-3.} 
The light curves of the source around the dates of Main (panels A-C) and ToO (panels D and E) observations, as obtained with various telescopes.  IXPE dates are marked with blue stripes.
Note high intraday variations of the radio flux caused by the orbital variability.
Data are given as the mean values with error bars corresponding to their variance.}\label{fig:radio_lc}
\end{figure}

\begin{figure}
\centering
\includegraphics[width=0.5\textwidth]{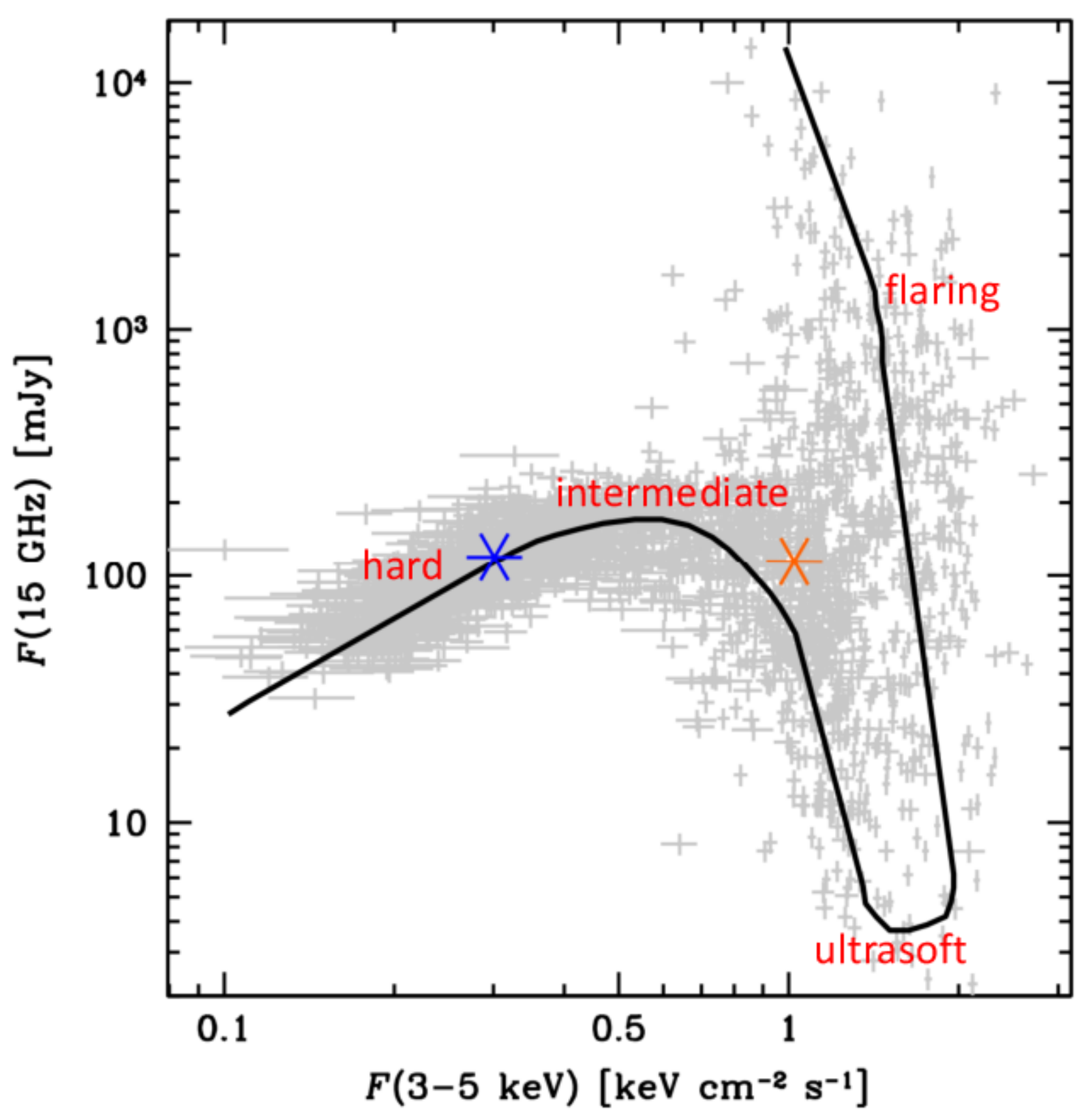} 
\caption{\textbf{Radio-X-ray evolution track from historical radio and X-ray observations.} 
Grey points constitute data analysed in \cite{Zdziarski2016}. Spectral states are indicated with red. Blue and orange stars indicate the fluxes during the Main and ToO observations, respectively.}\label{fig:diagram} 
\end{figure}

\begin{figure}
\centering
\includegraphics[width=0.9\textwidth]{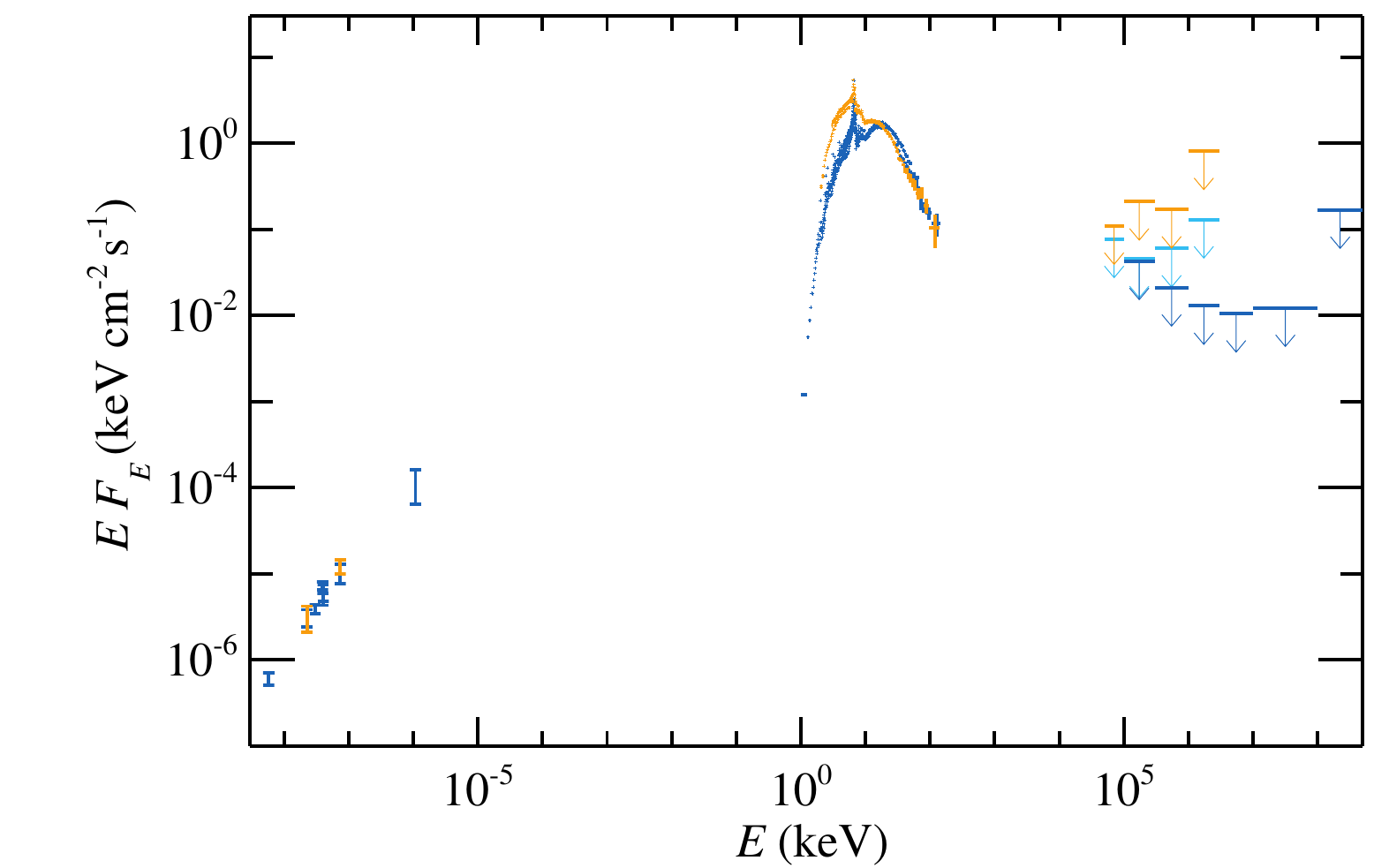} 
\caption{\textbf{Broadband spectral energy distribution of Cyg X-3.} 
The SED for the Main (blue) and ToO (orange) observations are from the facilities described in the text. Error bars correspond to 1$\sigma$ levels.}\label{fig:broadband_spectra} 
\end{figure}

\begin{figure}
\centering
\includegraphics[width=0.9\textwidth]{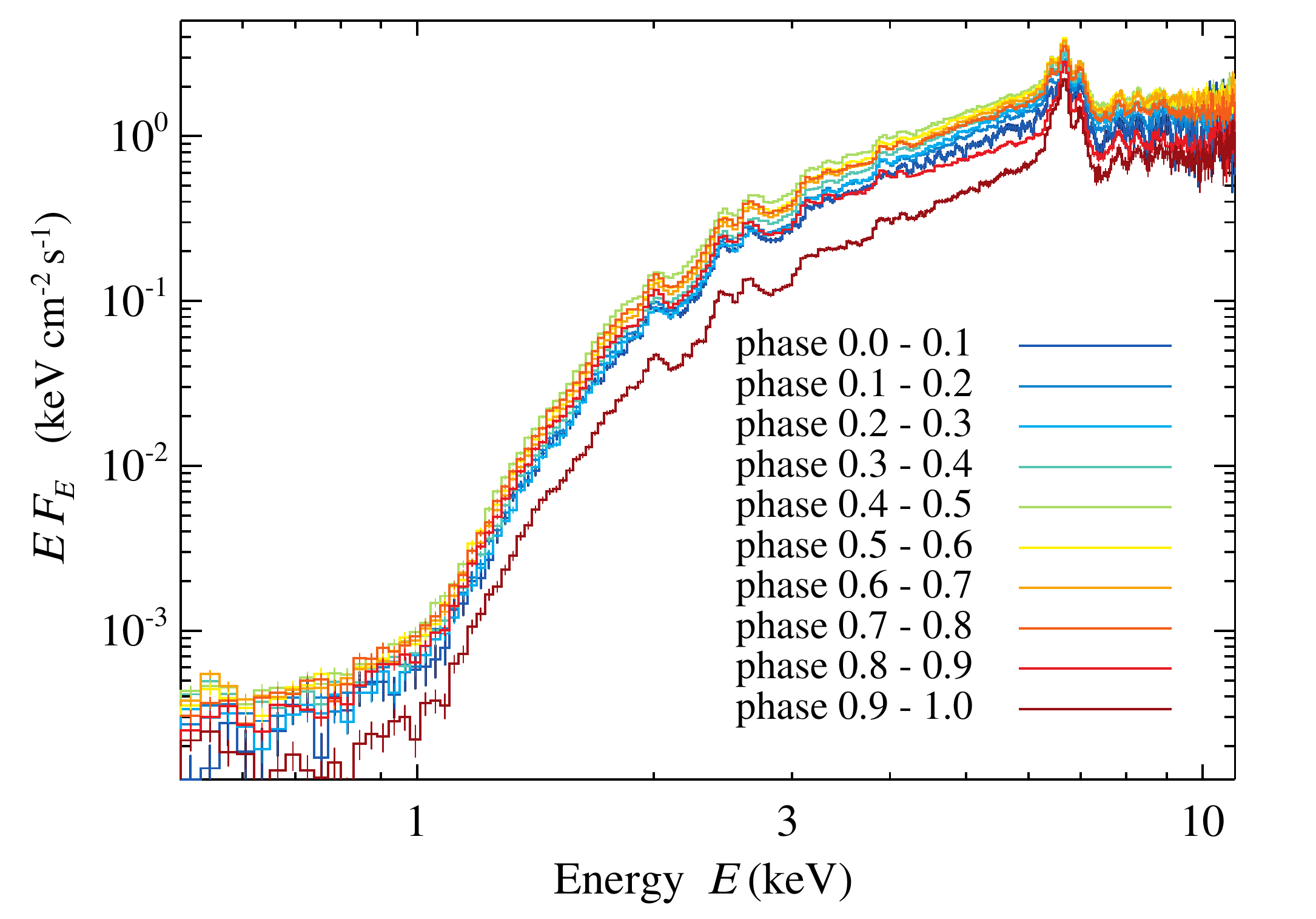}
\caption{\textbf{X-ray SED of Cyg X-3 from NICER.}
Orbital phase-folded X-ray spectra are taken during the contemporaneous observations in the Main run. From different phase intervals are presented in different colours.}
\label{fig:nicer_orbit_folded}
\end{figure}

\begin{figure}%
\centering
\includegraphics[width=0.9\textwidth]{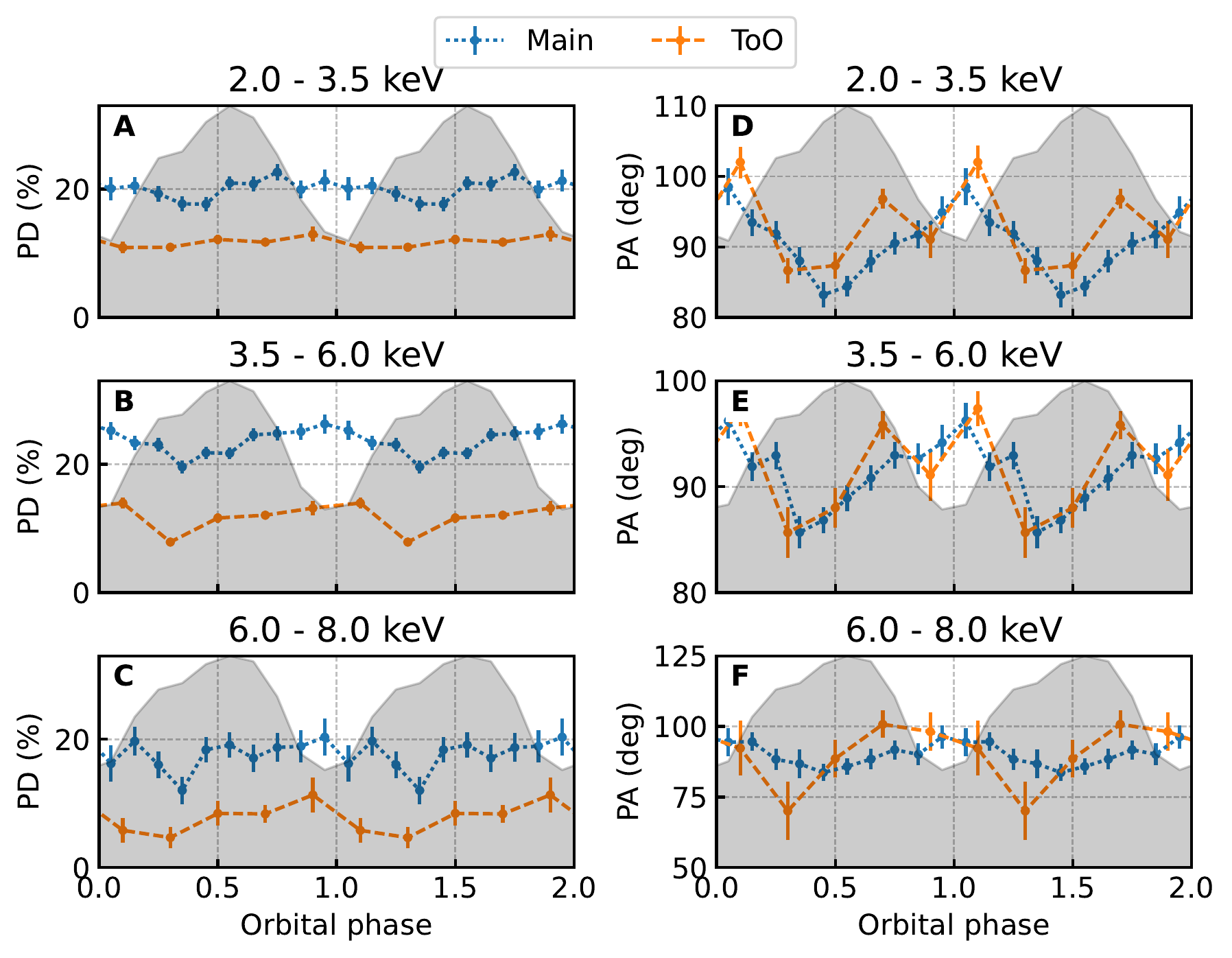}
\caption{\textbf{Orbital phase dependence of polarization.}  
The PD (A)--(C) and PA (D)--(F) in different energy bands (2--3.5~keV, A, D; 3.5--6~keV, B, E; 6--8~keV, C, F) for the Main (in blue) and ToO (in orange) observations are shown. Orbital profiles of IXPE flux are shown in each panel as shaded areas. Error bars correspond to 1$\sigma$ uncertainty level.}\label{fig:pol_orb_energy_all}
\end{figure}

\begin{figure} 
\centering
\includegraphics[width=\textwidth]{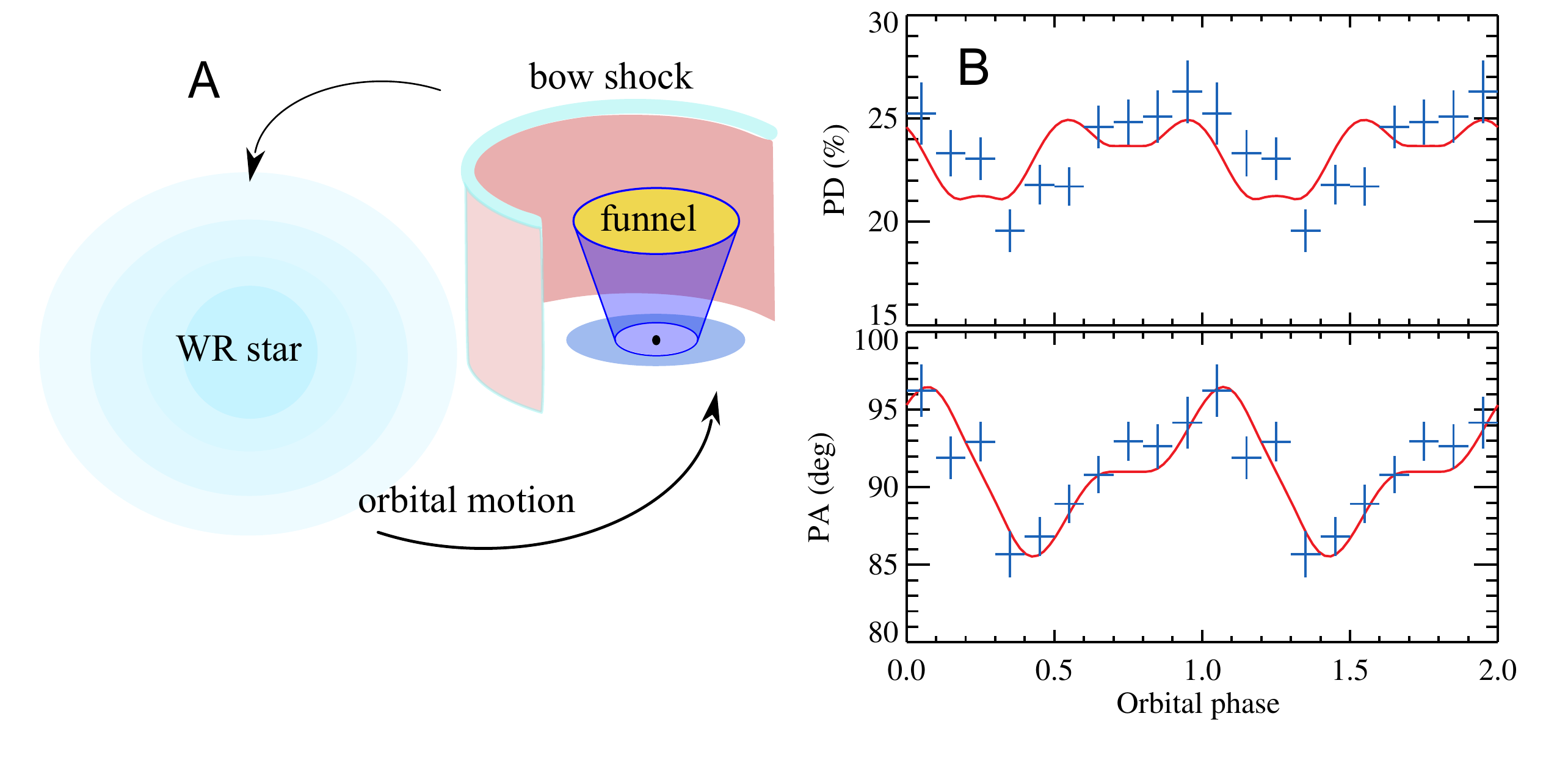}
\caption{\textbf{Modelling orbital variations of the PD and PA.} 
\textbf{(A)} Geometry of the reflector. 
\textbf{(B)} Dependence of the PD and PA in the 3.5--6~keV band on orbital phase for the Main observation is shown with blue crosses.  The red curve is the model of the reflection from a bow shock. 
}\label{fig:bow_shock}
\end{figure}

\begin{figure}
\centering
\includegraphics[width=\textwidth]{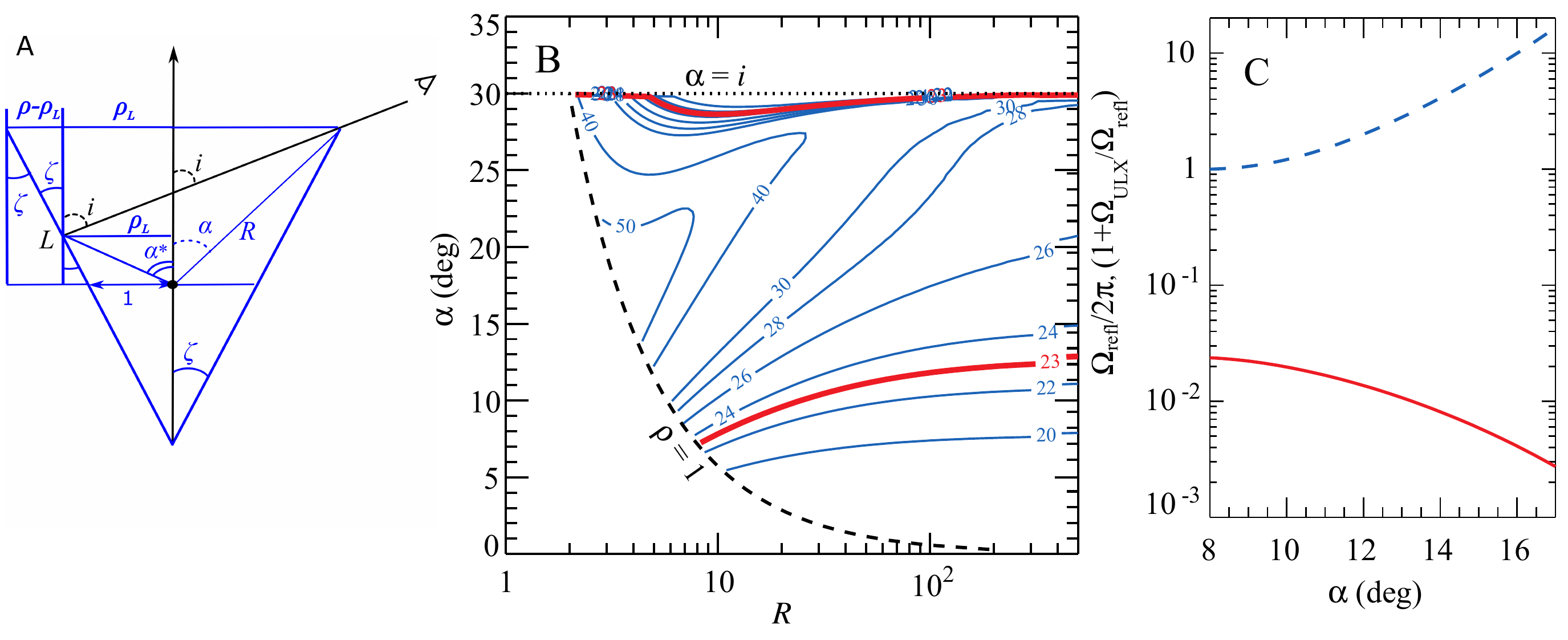}
\caption{\textbf{Detailed geometry of the reflecting funnel, its polarimetric characteristics, reflection and amplification factors.}
(A) Geometry of the funnel is shown with $L$ being the lowest visible point for the given inclination $i$, and the angle $\alpha^*$ is its colatitude.  
(B) The contour plots of constant PD (in \%) for the fixed observer inclination ($i=30\degr$), as function of the model parameters ($\alpha$, $R$). 
The region above $\alpha=i$ is not allowed because the central source would be visible. 
The region below $\rho=1$ curve (i.e. $R=1/\sin\alpha$) corresponds to an outflow converging towards the axis, which is not possible.  
Red contours show the allowed model parameters.
(C) Dependence of the solid angle of the reflecting surface ($\Omega_{\rm refl}/2\pi$, red solid curve) and the factor determining the intrinsic luminosity ($\displaystyle 1 + {\Omega_{\rm ULX}}/{\Omega_{\rm refl}}$, blue dashed curve) on the angle $\alpha$.
}\label{fig:model_alpha_rho} 
\end{figure}

\begin{figure}
\centering
\includegraphics[width=\textwidth]{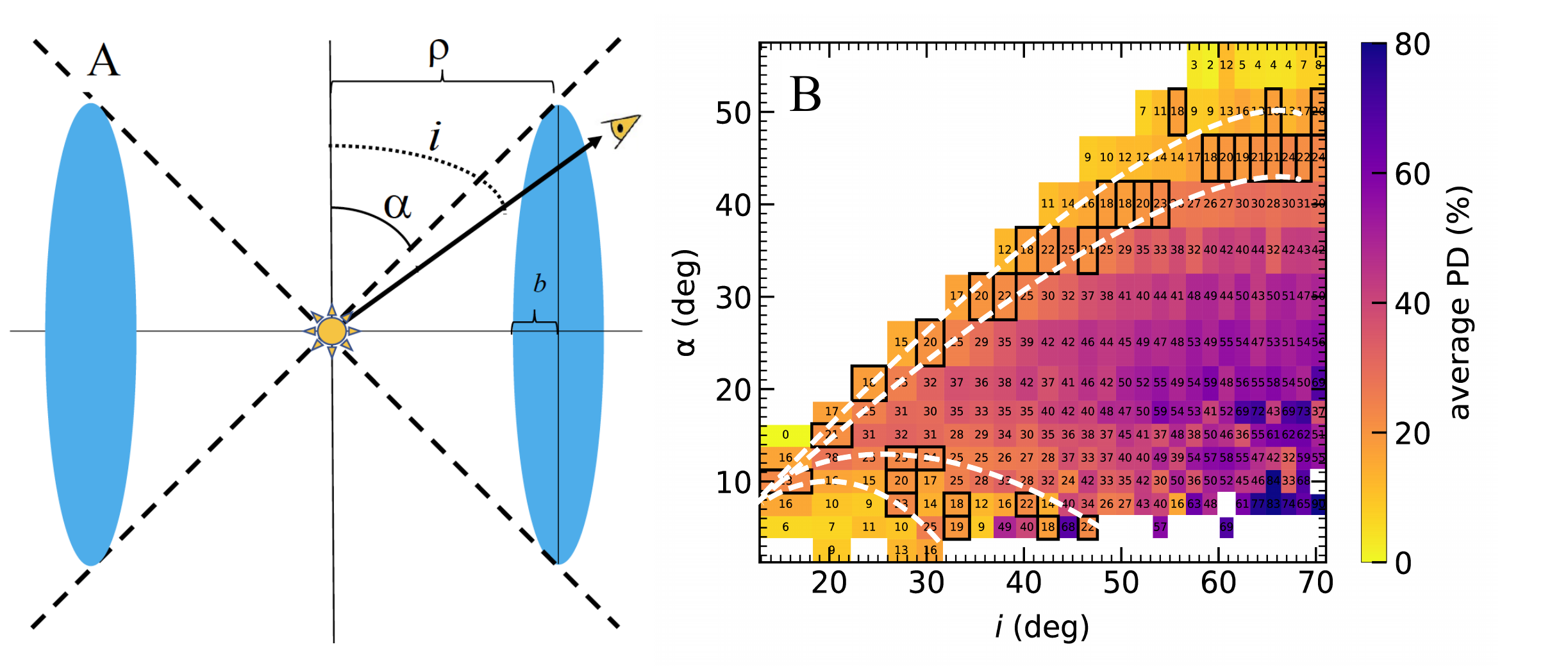}
\caption{\textbf{Results of Monte-Carlo simulations.}
\textbf{(A)} The geometry of the reflector (elliptical torus in blue) and main parameters of the funnel explored by the Monte-Carlo modelling.
\textbf{(B)} The simulated 2--8 keV PD versus observer's inclination and half-opening angle of the torus for $b = \rho/4$, $\tau_{\rm e} = 7$ and $N_{\rm He} = 8.5 \times 10^{23}$cm$^{-2}$ (the same display as in Fig.~\ref{fig:model_funnel} for the analytical model). The black rectangles and white dashed lines mark the regions where the reprocessed component gives PD$=21\pm3$\%.
} \label{fig:MC_geometries}
\end{figure}

\clearpage
\newpage
\section*{Tables}

\begin{table}[h]
\caption{Summary of contemporaneous X-ray and $\gamma$-ray observations. 
}\label{tab:mw_xray}
\begin{tabular}{@{}lccc@{}}
\toprule
Facility & Energy (keV) & MJD$-$59800  & Average flux (keV~cm$^{-2}$~s$^{-1})$ \\
\midrule
IXPE      & 2--8     & 66--71, 83--89  & 0.96 \\
              &          & 138--142  & 2.6 \\
NICER     & 0.5--12  & 84--87 & 1.6	  \\
ART-XC    & 4--30    & 87   & 2.9    \\
INTEGRAL  & 20--100   & 84--88  & $1.1$	  \\
         &           & 138  & $0.9$	  \\   
NuSTAR    & 3--50    & 65--66  & 3.5	  \\
                   & & 138--139     & 6.0 \\
AGILE     & $10^5-5\times10^{7}$   & 66--71, 83--89  & $<0.033$ 	  \\
      &     & 138--142  & $<0.22$ 	  \\
Fermi     & $10^{5}-10^{8}$   & 62--73  & $\lesssim 0.01$	  \\
\botrule
\end{tabular}
\end{table}

\begin{table}[h]
\caption{Summary of radio and sub-millimeter  observations. IXPE observations were performed on MJD~59866--59871, 59883--59889 and 59938--59942.}\label{tab:mw_radio}
\begin{tabular}{@{}lcccccc@{}}
\toprule
Telescope & Date & Frequency  & Average flux & Variance & PD & PA  \\
         &  MJD$-$59800 & (GHz)  & (mJy) & (mJy) & \%   & (deg) \\
\midrule
SMA           &  71  & 225  &  76  &  36  & $2.8\pm1.1$ & $-28\pm12$ \\
              &  85  & 225  &  86 & 35  & $2.2\pm0.4$ & $-6\pm6$ \\
AMI-LA      &   63--90  &   15.5  &   106 & 27 &&	\\
            &    137--139 & 15.5  &   126 & 24 && \\
Medicina    &   66--70  &   8.4 &   118 & 26	&& \\
Effelsberg &    61--70 &    8.3 &  99 & 16  &&\\
Effelsberg &    61--70 &    6.3 &  99 & 12 &&\\
RATAN      &   84--88  &   8.2 &   142 & 15 &&\\
      &      &   4.7 &   106 & 24 &&\\
      &   138   &   4.7 &   107 & 36 &&\\
uGMRT   &   85--86  &   1.2 &   81  &  14 && \\
\botrule
\end{tabular} 
\end{table}

\begin{table}[h]
\caption{Model parameters for the spectro-polarimetric fit of the IXPE and NuSTAR data. The model is: {\sc constant $\times$ mpl $\times$ (polconst $\times$ reflect $\times$ smedge $\times$ nthComp + polconst $\times$ gaussian)}. Uncertainties are calculated with the \textsc{XSPEC/error} command at 90\% confidence level.
} \label{tab:specpol}
\begin{tabular}{@{}lcc@{}}
\toprule
Parameter   & Main & ToO \\
\midrule
polconst\_refl A & 0.228$^{+0.008}_{-0.005}$ & 0.100$^{+0.007}_{-0.008}$ \\ 
polconst\_refl psi (deg) & 90.9$^{+0.6}_{-1.0}$ & 90.6$^{+2.5}_{-2.3}$ \\ 
reflect rel\_refl & $-1.0$ (frozen) & $-1.0$ (frozen) \\ 
reflect Redshift & 0.0 (frozen) & 0.0 (frozen) \\ 
reflect abund & 1.0 (frozen) & 1.0 (frozen) \\ 
reflect Fe\_abund & 0.69$^{+0.02}_{-0.04}$ & 0.47$^{+0.02}_{-0.06}$ \\ 
reflect cosIncl & 0.23$^{+0.11}_{-0.05}$ & 0.051$^{+0.013}_{-0.001}$ \\ 
smedge edgeE (keV) & 8.66$^{+0.08}_{-0.15}$ & 8.83$^{+0.08}_{-0.04}$ \\ 
smedge MaxTau & 0.48$^{+0.13}_{-0.03}$ & 0.35$^{+0.02}_{-0.06}$ \\ 
smedge index & $-2.7$ (frozen) & $-2.7$ (frozen) \\ 
smedge width & 1.8$^{+0.9}_{-0.3}$ & 0.49$^{+0.05}_{-0.11}$ \\ 
diskpbb $kT_{\rm in}$ (keV) & -- & 1.10$^{+0.02}_{-0.03}$ \\ 
diskpbb p & -- & 0.50$^{+0.05}_{-0.50}$ \\ 
diskpbb norm ($\times10^3$) & -- & 7.9$^{+0.9}_{-1.8}$ \\ 
nthComp Gamma & 2.62$^{+0.07}_{-0.03}$ & 2.98$^{+0.02}_{-0.06}$ \\ 
nthComp $kT_{\rm e}$ (keV) & 30$^{+16}_{-5}$ & 47$^{+6}_{-11}$ \\ 
nthComp $kT_{\rm bb}$ (keV) & 0.62$^{+0.03}_{-0.02}$ & 1.21$^{+0.05}_{-0.20}$ \\ 
nthComp norm & 3.6$^{+0.7}_{-0.8}$ & 2.5$^{+1.3}_{-0.3}$ \\ 
polconst\_gauss A & 0.00$^{+0.02}$ & 0.00$^{+0.06}$ \\ 
polconst\_gauss psi (deg) & 91 (linked) & 91 (linked) \\ 
gaussian LineE (keV) & 6.64$^{+0.01}_{-0.01}$ & 6.53$^{+0.02}_{-0.02}$ \\ 
gaussian Sigma (keV) & 0.261$^{+0.014}_{-0.013}$ & 0.19$^{+0.03}_{-0.02}$ \\ 
gaussian norm ($\times10^{-3}$) & 12.76$^{+0.44}_{-0.37}$ & 17.1$^{+1.7}_{-1.1}$ \\IXPE/det1\_constant factor & \multicolumn{2}{c}{1.76$^{+0.11}_{-0.06}$}\\ 
IXPE/det2\_constant factor & \multicolumn{2}{c}{$1.62^{+0.10}_{-0.06}$} \\ 
IXPE/det3\_constant factor & \multicolumn{2}{c}{1.63$^{+0.10}_{-0.06}$} \\
IXPE\_mpl gamma & 
\multicolumn{2}{c}{$0.29^{+0.04}_{-0.02}$}\\ 
IXPE/det1 gain slope & $0.9990^{+0.0012}_{-0.0005}$ & -- \\ 
IXPE/det1 gain offset (keV) & $-0.054^{+0.006}_{-0.004}$ & -- \\ 
IXPE/det2 gain slope & $0.9790^{+0.0004}_{-0.0012}$ & -- \\ 
IXPE/det2 gain offset (keV) & $0.083^{+0.004}_{-0.007}$ & -- \\ 
IXPE/det3 gain slope & $0.9948^{+0.0006}_{-0.0013}$ & -- \\ 
IXPE/det3 gain offset (keV) & $-0.031^{+0.004}_{-0.007}$ & -- \\ 
NuSTAR gain slope & 1.0 (frozen) & -- \\ 
NuSTAR gain offset (keV) & $-0.103^{+0.007}_{-0.001}$ & -- \\ 
$\chi^2/$d.o.f. & 935.77/881 & 921.63/885 \\
\botrule
\end{tabular}
\end{table}

\clearpage

\section*{Data availability}

The IXPE, Nustar, NICER, INTEGRAL and Fermi data are freely available in the HEASARC IXPE Data Archive (\url{https://heasarc.gsfc.nasa.gov}). 
The SRG ART-XC data are available via \url{ftp://hea.iki.rssi.ru/public/SRG/ART-XC/data/Cyg_X-3/artxc_cygx3_04-20keV_lcurve.qdp}.
The multiwavelength data are available on request from the individual observatories.

\section*{Code availability}

The analysis and simulation software \textsc{ixpeobssim} developed by IXPE collaboration and its documentation is available publicly through the web-page \url{https://ixpeobssim.readthedocs.io/en/latest/?badge=latest.494}.
\textsc{xspec} is distributed and maintained under the aegis of the HEASARC and can be downloaded as part of HEAsoft from
\url{http://heasarc.gsfc.nasa.gov/docs/software/lheasoft/download.html}.
MIR software package for SMA data: \url{https://lweb.cfa.harvard.edu/~cqi/mircook.html}.
The {\tt STOKES} code version \textit{v2.07} is available upon reasonable request from the authors.

\section*{Acknowledgements}

The Imaging X-ray Polarimetry Explorer (IXPE) is a joint US and Italian mission.
The US contribution is supported by the National Aeronautics and Space Administration (NASA) and led and managed by its Marshall Space Flight Center (MSFC), with industry partner Ball Aerospace (contract NNM15AA18C). 
The Italian contribution is supported by the Italian Space Agency (Agenzia Spaziale Italiana, ASI) through contract ASI-OHBI-2017-12-I.0, agreements ASI-INAF-2017-12-H0 and ASI-INFN-2017.13-H0, and its Space Science Data Center (SSDC), and by the Istituto Nazionale di Astrofisica (INAF) and the Istituto Nazionale di Fisica Nucleare (INFN) in Italy.
For the AMI observations we thank the staff of the Mullard Radio Astronomy Observatory, University of Cambridge, for their support in the maintenance, and operation of telescope, and we acknowledge support from the European Research Council under grant ERC-2012-StG-307215 LODESTONE.
The Submillimeter Array (SMA) is a joint project between the Smithsonian Astrophysical Observatory and the Academia Sinica Institute of Astronomy and Astrophysics, and is funded by the Smithsonian Institution and the Academia Sinica.
SMA is operated on Maunakea which is a culturally important site for the indigenous Hawaiian people; we are privileged to study the cosmos from its summit. 
Partly based on observations with the 100-m telescope of the MPIfR (Max-Planck-Institut fuer Radioastronomie) at Effelsberg. 
The research leading to these results has received funding from the European Union's Horizon 2020 research and innovation programme under grant agreement No 101004719 (ORP).
The AGILE Mission is funded by the Italian Space Agency (ASI) with scientific and programmatic participation by the Italian National Institute for Astrophysics (INAF) and the Italian National Institute for Nuclear Physics (INFN).
This investigation was supported by the ASI grant I/028/12/7-2022.

The authors thank Hua Feng for providing the data on the representative ULX models.
F.Mu, A.D.M., F.L.M., E.Co., P.So., S.F. and R.F., are partially supported by MAECI with grant CN24GR08 ``GRBAXP: Guangxi-Rome Bilateral Agreement for X-ray Polarimetry in Astrophysics''.
A.V., Ju.Pou. and S.S.T. acknowledge support from the Academy of Finland (grants 333112, 347003, 349144, 349373, 349906, 355672). 
A.A.M. is supported by the UKRI Stephen Hawking fellowship.
H.K. and N.R.C. acknowledge NASA support under grants 80NSSC18K0264, 80NSSC22K1291, 80NSSC21K1817, and NNX16AC42G.
V.D. thanks the German Academic Exchange Service (DAAD) for the travel grant 57525212. 
A.I. acknowledges support from the Royal Society.
Ja.Pod., M.D., J.S. and V.K. thank for the support from the GACR project 21-06825X and the institutional support from RVO:67985815.
We thank the staff of the GMRT that made these observations possible. GMRT is run by the National Centre for Radio Astrophysics of the Tata Institute of Fundamental Research.
R.K. acknowledges the support of the Department of Atomic Energy, Government of India, under project no. 12-R\&D-TFR-5.02-0700. 
M.M. is supported by NASA contract NAS8-03060.
S.A.T. is supported by the Ministry of Science and Higher Education of the Russian Federation grant 075-15-2022-262 (13.MNPMU.21.0003).
A.A.Z. acknowledges support from the Polish National Science Center under the grant 2019/35/B/ST9/03944.

\section*{Author contributions}
A.V. led the modelling of the data and the writing of the paper. 
F.Mu. led the analysis of the IXPE data.
Ju.Pou. led the analytical modelling and contributed to writing of the paper. 
Ja.Pod. performed Monte-Carlo simulations supporting the modelling. 
M.D. led the work of the IXPE Topical Working Group on Accreting stellar-mass black holes. 
A.D.R., E.Ch., P.K. and R.A.S.  contributed with parts of the paper and its content.
F.C., A.D.M, S.F., H.K., F.L.M., A.A.L., S.V.M., A.R., N.R.C., J.F.S., S.S.T., A.A.Z. and J.J.E.K., I.A.M., G.P. and C.P. contributed to planning, reducing and analysing the X-ray/gamma-ray data.
V.L., A.A.M. and D.M. contributed to analytical estimates and modelling.
J.S.B., N.Bur., E.E., D.A.G., M.G., R.K., A.K., M.McC., N.N., M.Pi., R.R., S.R., A.S., J.S., S.T. and P.T. contributed with radio/sub-millimeter data.
S.B., E.Co., J.A.G., A.I., F.Mar., G.M., P.So., F.To., F.U., M.C.W. and K.W. contributed with discussion of methods and conclusions.
The remaining authors have contributed to the design, science case of the IXPE mission and planning of observations relevant to the present paper.   
%
%
%
%


\end{document}